\documentclass[onecolumn,pra,showpacs,superscriptaddress,preprintnumbers]{revtex4}

\usepackage{graphicx}
\usepackage{bm}
\usepackage{amsmath,amssymb,epsfig}
\usepackage{color}

\newcommand{\mathd}{\mathrm{d}}

\newcommand{\tmmathbf}[1]{\ensuremath{\boldsymbol{#1}}}

\newcommand{\tmem}[1]{{\em #1\/}}

\newcommand{\tmop}[1]{\ensuremath{\text{#1}}}
\newcommand{\bignone}{}

\begin{document}

\title{Stochastic simulation algorithm for the quantum linear Boltzmann equation}

\author{Marc Busse}
\affiliation{Arnold Sommerfeld Center for Theoretical Physics,
Ludwig-Maximilians-Universit{\"a}t M{\"u}nchen, Theresienstra{\ss}e 37,
80333 Munich, Germany
}
\author{Piotr Pietrulewicz}
\affiliation{Arnold Sommerfeld Center for Theoretical Physics,
Ludwig-Maximilians-Universit{\"a}t M{\"u}nchen, Theresienstra{\ss}e 37,
80333 Munich, Germany
}
\author{Heinz-Peter Breuer}
\affiliation{Physikalisches Institut,
Universit{\"a}t Freiburg, Hermann-Herder-Stra{\ss}e 3,
79104 Freiburg, Germany
}
\author{Klaus Hornberger}
\affiliation{Arnold Sommerfeld Center for Theoretical Physics,
Ludwig-Maximilians-Universit{\"a}t M{\"u}nchen, Theresienstra{\ss}e 37,
80333 Munich, Germany
}
\affiliation{Max Planck Institute for the Physics of Complex Systems,
N{\"o}thnitzer Stra{\ss}e 38,
01187 Dresden, Germany
}

\date{May 3, 2010}

\begin{abstract}
We develop a Monte Carlo wave function algorithm for the quantum linear Boltzmann equation, a Markovian master equation describing the quantum motion of a test particle interacting with the particles of an environmental background gas. The algorithm leads to a numerically efficient stochastic simulation procedure for the most general form of this integro-differential equation, which involves a five-dimensional integral over  microscopically defined scattering amplitudes that account for the gas interactions in a non-perturbative fashion. The simulation technique is used to assess various limiting forms of the quantum linear Boltzmann equation, such as the limits of pure collisional decoherence and quantum Brownian motion, the Born approximation and the classical limit.
Moreover, we extend the method to allow for the simulation of the dissipative and decohering dynamics of superpositions of spatially localized wave packets, which enables the study of many physically relevant quantum phenomena, occurring e.g. in the interferometry of massive particles.
\end{abstract}


\preprint{\textsf{published in Phys.~Rev.~E~{82}, 026706 (2010)}}

\maketitle

\section{Introduction \label{sec:introduction}}

The motion of a quantum particle interacting with a surroundings particle gas is characterized by collision-induced decoherence as well as dissipation and thermalization effects. An appropriate master equation which provides a unified quantitative description of both phenomena in a mathematically consistent way is the quantum linear Boltzmann equation (QLBE), proposed in its weak-coupling form in \cite{Vacchini2001b,Vacchini2002a}, and in final form in {\cite{Hornberger2006b}}. This equation represents the quantum mechanical generalization of the classical linear Boltzmann equation which describes the motion of a distinguished test particle under the influence of elastic collisions with an ideal, stationary background gas. The QLBE may be derived on the basis of a monitoring approach \cite{Hornberger2007b} which permits a non-perturbative treatment of the interactions with the environmental gas particles \cite{Hornberger2006b,Hornberger2008a}. These interactions may therefore be strong and the test particle may be in a state which is far from equilibrium. A condition for the applicability of the monitoring approach is that three-particle collisions are sufficiently unlikely, and that successive collisions of the test particle with the same gas particle are negligible on the the relevant time scale. These conditions are fulfilled in the case of an ideal background gas in a stationary equilibrium state. A further condition is that the interactions are short-ranged so that scattering theory may be applied.

The mathematical structure of the QLBE is rather involved and analytical solutions of this equation are known only for some specific limiting cases \cite{Vacchini2009a}. Moreover, the spatially nonlocal structure of this equation makes a direct numerical integration through deterministic methods extremely demanding. However, being in Lindblad form, the QLBE allows one to apply the Monte Carlo wave function techniques 
\nocite{Gardiner1992a}
\nocite{Carmichael1993a}
\nocite{Molmer1993a}
\nocite{Molmer1996a}
\cite{Gardiner1992a, Carmichael1993a, Molmer1993a, Molmer1996a, Breuer2007b}. As has been demonstrated in Ref.~\cite{Breuer2007a} these techniques lead to a simple and numerically efficient stochastic simulation method in the momentum representation of the test particle's density matrix, employing the translational covariance of the QLBE.

The simulation technique developed in \cite{Breuer2007a} is restricted to the QLBE within the Born approximation \cite{Vacchini2001b,Vacchini2002a}, in which the scattering cross section depends only on the momentum transfer of the scattered particles, yielding a considerably simplified equation of motion. Here we generalize this stochastic approach to the full QLBE allowing for an arbitrary form of the microscopic interaction between the test particle and the ambient gas particles and, thus, arbitrary scattering amplitudes. In addition, we show how the algorithm can be extended to simulate efficiently the dynamics of spatially localized wave packets. This enables the exact numerical treatment of many physically relevant phenomena, such as the loss of coherence in position space and the determination of the fringe visibility in interferometric devices, as well as the assessment of the quality of various approximations of the QLBE.

The paper is organized as follows. Section \ref{sec:QLBE} contains a brief account of the QLBE and summarizes the most important limiting forms of this equation. In Sec.~\ref{sec:MCU} we develop the Monte Carlo simulation algorithm for the full three-dimensional QLBE in momentum space. Our numerical simulation results are presented in Sec.~\ref{sec:Numericalresults}. We discuss examples for the decoherence of superpositions of momentum eigenstates, the loss of coherence of superpositions of spatially localized wave packets, the decohering influence of the background gas on the fringe visibility of interference experiments, relaxation and thermalization processes, and the diffusion limit. Finally, Sec.~\ref{sec:conclusion} contains a brief summary of the results and our conclusions.

\section{The quantum linear Boltzmann equation \label{sec:QLBE}}

\subsection{General form of the master equation}

The quantum linear Boltzmann equation (QLBE) is a Markovian master equation for the reduced density operator $\rho$ describing the evolution of a test particle in an ideal gas environment. It has the form $\dot{\rho} =\mathcal{L} \rho$, where the generator is of Lindblad structure,
\begin{eqnarray}
  \mathcal{L} \rho & = & \frac{1}{i \hbar} \left[ \frac{\mathsf{P}^2}{2 M} +
  {H}_n \left( \mathsf{\mathsf{P}} \right), \rho \right] +\mathcal{D}
  \rho \, . \bignone  \label{eq:QLBELind}
\end{eqnarray}
Here ${H}_n \left( \mathsf{\mathsf{P}} \right)$ describes the energy
shift due to the interaction with the background gas; it will be neglected in
the following, since it is usually small. The incoherent part of the
interaction is accounted for by the superoperator $\mathcal{D}$, which can be
expressed as {\cite{Hornberger2006b,Hornberger2008a,Vacchini2009a}}
\begin{eqnarray}
  \mathcal{D} \rho & = & \int_{\mathbb{R}^3} \mathd \tmmathbf{Q}
  \int_{\tmmathbf{Q}^{\perp}} \mathd {\tmmathbf{k}_{\bot}} \Big( e^{i\tmmathbf{Q}
  \cdot \mathsf{X} / \hbar} L \left( {\tmmathbf{k}_{\bot}}, \mathsf{P}, \tmmathbf{Q}
  \right) \rho L^{\dag} \left( {\tmmathbf{k}_{\bot}}, \mathsf{P}, \tmmathbf{Q} \right)
  e^{- i\tmmathbf{Q} \cdot \mathsf{X} / \hbar}  \nonumber\\
  &  &  - \frac{1}{2}  \left\{ \rho, L^{\dag} \left( {\tmmathbf{k}_{\bot}},
  \mathsf{P}, \tmmathbf{Q} \right) L \left( {\tmmathbf{k}_{\bot}}, \mathsf{P},
  \tmmathbf{Q} \right) \right\} \bignone \Big) \bignone \bignone \,,
  \label{eq:QLBE}
\end{eqnarray}
with $\mathsf{X} = \left( \mathsf{X}_1, \mathsf{X}_2, \mathsf{X}_3 \right)$ the
position and $\mathsf{P} = \left( \mathsf{P}_1, \mathsf{P}_2, \mathsf{P}_3 \right)$  the momentum operator of the test particle. The
integration variables are given by $\tmmathbf{Q}$, the momentum transfer experienced in a single collision, and ${\tmmathbf{k}_{\bot}}$, corresponding to the momentum of a gas particle. The ${\tmmathbf{k}_{\bot}}$-integration is carried out over the plane $\tmmathbf{Q}^{\perp} = \left\{ {\tmmathbf{k}_{\bot}} \in \mathbb{R}^3
|{\tmmathbf{k}_{\bot}} \cdot \tmmathbf{Q}= 0 \right\}$ perpendicular to the momentum transfer $\tmmathbf{Q}$.

The operator-valued function $L \left( {\tmmathbf{k}_{\bot}}, \mathsf{P}, \tmmathbf{Q}
\right)$ contains all the details of the collisional interaction with the gas;
these are the gas density $n_{\tmop{gas}}$, the momentum distribution function
$\mu \left( {\tmmathbf{p}} \right)$ of the gas, and the elastic scattering
amplitude $f \left( \tmmathbf{p}_f, \tmmathbf{p}_i \right)$. It is defined by
{\cite{Hornberger2006b,Hornberger2008a,Vacchini2009a}} 
\begin{eqnarray}
  L \left( {\tmmathbf{k}_{\bot}}, \mathbf{P}, \tmmathbf{Q} \right) & = &
  \sqrt{\frac{n_{\tmop{gas}} m}{m_{\ast}^2 Q}} f \left( \mathbf{p}_{\rm rel} \left(
  {\tmmathbf{k}_{\bot}}, \mathbf{P}_{\perp \tmmathbf{Q}} \right) -
  \frac{\tmmathbf{Q}}{2}, \mathbf{p}_{\rm rel} \left( {\tmmathbf{k}_{\bot}},
  \mathbf{P}_{\perp \tmmathbf{Q}} \right) + \frac{\tmmathbf{Q}}{2} \right)
  \nonumber\\
  &  & \times \sqrt{\mu \left( {\tmmathbf{k}_{\bot}} +
  \frac{m}{m_{\ast}} \frac{\tmmathbf{Q}}{2_{}} + \frac{m}{M}
  \mathbf{P}_{\|\tmmathbf{Q}} \right)} \, .  \label{eq:QLBEL}
\end{eqnarray}
Here $m_{\ast} \equiv mM / \left( m + M \right)$ is the reduced mass, $Q
\equiv \left| \tmmathbf{Q} \right|$ gives the modulus of the momentum
transfer $\tmmathbf{Q}$, and the function
\begin{eqnarray}
  \mathbf{p}_{\rm rel} \left( {\tmmathbf{p}}, \mathbf{P} \right) & \equiv &
  \frac{m_{\ast}}{m} {\tmmathbf{p}}- \frac{m_{\ast}}{M} \mathbf{P} \,,
  \label{eq:rel}
\end{eqnarray}
defines relative momenta. The subscripts $\|\tmmathbf{Q}$ and $\perp \tmmathbf{Q}$
denote the parts of a given vector $\tmmathbf{P}$ parallel and perpendicular to $\tmmathbf{Q}$, i.e.
\begin{eqnarray}
  \tmmathbf{P}_{\|\tmmathbf{Q}} & = & \frac{\left( \tmmathbf{P} \cdot
  \tmmathbf{Q} \right) \tmmathbf{Q}}{Q^2} \,,  \label{eq:comppar}\\
  \tmmathbf{P}_{\perp \tmmathbf{Q}} & = &
  \tmmathbf{P}-\tmmathbf{P}_{\|\tmmathbf{Q}} \, .  \label{eq:compperp}
\end{eqnarray}
We note that the QLBE described by Eqs.~(\ref{eq:QLBELind}) and (\ref{eq:QLBE}) has the structure of a translation-covariant master equation in Lindblad form, according to the general characterization given by Holevo
{\cite{Holevo1993a,Holevo1993b,Holevo1995a,Holevo1996a,Petruccione2005a}}.
This feature will be important below when applying the stochastic unraveling of the QLBE.

\subsection{Limiting forms \label{sec:Limitingforms}}

Suitable limiting procedures reduce the QLBE to other well-known evolution equations, whose solutions are (at least partly) understood. These relations allow us to interpret the numerical solutions of the QLBE later on. At the same time, the stochastic simulation technique of the full QLBE permits us to study the range of validity of these approximate evolution equations.

\subsubsection{Classical linear Boltzmann equation \label{sec:CLBE}}

To establish the connection to the classical linear Boltzmann equation one may consider the evolution of the diagonal elements $w \left( \tmmathbf{P} \right)
\equiv \langle \tmmathbf{P}| {\rho} |\tmmathbf{P} \rangle$ in the
momentum basis. As is shown in
{\cite{Hornberger2006b,Hornberger2008a,Vacchini2009a}} the incoherent part
of the QLBE implies that
\begin{eqnarray}
  \partial_t w \left( \tmmathbf{P} \right) & = & \int \mathd \tmmathbf{Q}
  \bignone \left[ M^{\tmop{cl}} \left( \tmmathbf{P}-\tmmathbf{Q} \rightarrow
  \tmmathbf{P} \right) w \left( \tmmathbf{P}-\tmmathbf{Q} \right)
  - M^{\tmop{cl}} \left( \tmmathbf{P} \rightarrow \tmmathbf{P}+\tmmathbf{Q}
  \right) w \left( \tmmathbf{P} \right) \, \right], 
  \label{eq:CLBE}
\end{eqnarray}
where the transition rates $M^{\tmop{cl}}$ are given by
\begin{eqnarray}
  M^{\tmop{cl}} \left( \tmmathbf{P} \rightarrow \tmmathbf{P}+\tmmathbf{Q}
  \right) & = & \int_{\tmmathbf{Q}^{\perp}} \mathd {\tmmathbf{k}_{\bot}} \bignone
  \left| L \left( {\tmmathbf{k}_{\bot}}, \tmmathbf{P}, \tmmathbf{Q} \right) \right|^2
  \label{eq:W1}\\
  & = & \frac{n_{\tmop{gas}} m}{m_{\ast}^2 Q} \int_{\tmmathbf{Q}^{\perp}}
  \mathd {\tmmathbf{k}_{\bot}} \mu \left( {\tmmathbf{k}_{\bot}} +
  \frac{m}{m_{\ast}} \frac{\tmmathbf{Q}}{2_{}} + \frac{m}{M}
  \mathbf{P}_{\|\tmmathbf{Q}} \right) \nonumber\\
  &  & \times \sigma \left( \mathbf{p}_{\rm rel} \left( {\tmmathbf{k}_{\bot}}, \mathbf{P}_{\perp \tmmathbf{Q}} \right) -
  \frac{\tmmathbf{Q}}{2}, \mathbf{p}_{\rm rel} \left( {\tmmathbf{k}_{\bot}},
  \mathbf{P}_{\perp \tmmathbf{Q}} \right) + \frac{\tmmathbf{Q}}{2} \right) \,
  .  \label{eq:W2}
\end{eqnarray}
Here $\sigma \left( \tmmathbf{p}_f, \tmmathbf{p}_i \right) \equiv \left| f
\left( \tmmathbf{p}_f, \tmmathbf{p}_i \right) \right|^2$ denotes the quantum mechanical scattering cross section.

According to Refs.~{\cite{Hornberger2006b,Hornberger2008a,Vacchini2009a}},
Eqs.~(\ref{eq:CLBE}) and (\ref{eq:W2}) agree with the collisional part of the
classical linear Boltzmann equation \cite{Cercignani1975a}. In addition, it is argued in {\cite{Vacchini2009a}} that the solution
of the QLBE becomes asymptotically diagonal in the momentum basis for any initial
state $\rho_0$, that is
$\langle \tmmathbf{P}|e^{\mathcal{L}t} \rho_0 |\tmmathbf{P}'\neq\tmmathbf{P} \rangle \rightarrow
  0$ as $ t \rightarrow \infty $. 
It follows that the QLBE asymptotically approaches  the classical linear Boltzmann equation for the population dynamics in momentum space. This fact will be important below when analyzing the diffusive behavior exhibited by the numerical solution of the QLBE.

\subsubsection{Pure collisional decoherence \label{sec:PCD}}

The complexity of the QLBE reduces considerably if one assumes the test particle to be much heavier than the gas particles. By setting the mass ratio $m/M$ equal to zero the Lindblad operators in (\ref{eq:QLBE}) no longer depend on the momentum operator $\mathsf{P}$ of the tracer particle, so that the
${\tmmathbf{k}_{\bot}}$-integration in (\ref{eq:QLBE}) can be carried out
{\cite{Hornberger2008a,Vacchini2009a}}. The QLBE then turns into the master equation of pure collisional decoherence 
\nocite{Gallis1990a}

\cite{Gallis1990a,Hornberger2003b},
\begin{eqnarray}
  \frac{\mathd}{\mathd t} \rho & = & \frac{1}{i \hbar} \left[
  \frac{\mathsf{P}^2}{2 M}, \rho \right] + \Gamma_{\tmop{eff}} \int \mathd
  \bignone \tmmathbf{Q}G \left( \tmmathbf{Q} \right) \left( e^{i\tmmathbf{Q}
  \mathsf{X} / \hbar} \rho e^{- i \mathsf{\tmmathbf{Q}X} / \hbar} - \rho
  \right) \,,  \label{eq:Collisionaldecoherence2}
\end{eqnarray}
where $G \left( \tmmathbf{Q} \right)$ denotes the normalized momentum transfer
distribution and $\Gamma_{\tmop{eff}}$ is the collision rate of the
gas environment, defined by the thermal average
\begin{eqnarray}
  \Gamma_{\tmop{eff}} & \equiv & n_{\tmop{gas}} \int_{\mathbb{R}^3} \mathd
  {\tmmathbf{p}} \frac{p}{m} \bignone \mu \left( {\tmmathbf{p}} \right) \sigma
  \left( p \right) \, .  \label{eq:Geff}
\end{eqnarray}
It leads to a localization in position space as can be seen by neglecting the Hamiltonian part in (\ref{eq:Collisionaldecoherence2}) for large $M$. The solution then takes the form
\begin{eqnarray}
  \langle \tmmathbf{X} | {\rho} \left( t \right) | \tmmathbf{X'} \rangle
  & = & e^{- F \left( \tmmathbf{X} - \tmmathbf{X}' \right) t} \langle
  \tmmathbf{X} | {\rho} \left( 0 \right) | \tmmathbf{X'} \rangle \, .
  \label{eq:decayCD2}
\end{eqnarray}
The decay rate of spatial coherences is given by the {\tmem{localization
rate}} $F \left( \Delta \tmmathbf{X} \right) \geq 0$ which is related to the momentum transfer distribution $G \left( \tmmathbf{Q} \right)$ by
\begin{eqnarray}
  F \left( \tmmathbf{\tmmathbf{X} - \tmmathbf{X}'} \right) & = &
  \Gamma_{\tmop{eff}} \left[ 1 - \int \mathd \tmmathbf{Q} \, G \left(
  \tmmathbf{Q} \right) \exp \left( \frac{i}{\hbar} \tmmathbf{Q} \cdot \left(
  \tmmathbf{X} - \tmmathbf{X}' \right) \right) \bignone \right] \, .
  \label{eq:localizationfunctionCD}
\end{eqnarray}
The localization rate can be determined from the microscopic quantities as \cite{Vacchini2009a}
\begin{eqnarray}
  F \left( \tmmathbf{\tmmathbf{\tmmathbf{X} - \tmmathbf{X}'}} \right) & = &
  \Gamma_{\tmop{eff}} - 2 \pi n_{\tmop{gas}} \int_0^{\infty} \mathd v \mu
  \left( v \right) v \int_{- 1}^1 \mathd \cos \theta \, \bignone \, \left| f
  \left( \cos \theta ; E_{\tmop{kin}} \right) \right|^2 \bignone \hspace{1em}
  \nonumber\\
  &  & \times \tmop{sinc} \left[ 2 \sin \left( \frac{\theta}{2} \right)
  \frac{mv \left| \tmmathbf{\tmmathbf{\tmmathbf{X} - \tmmathbf{X}'}}
  \right|}{\hbar} \right] \,,  \label{eq:explicitformloc}
\end{eqnarray}
where $\theta$ denotes the scattering angle. Here we have assumed isotropic scattering,
so that $f \left( \tmmathbf{p}_f, \tmmathbf{p}_i \right) = f \left( \cos
\left( \tmmathbf{p}_f, \tmmathbf{p}_i \right) ; E_{\tmop{kin}} = p_i^2 / 2 m
\right)$. Equation (\ref{eq:explicitformloc}) will allow us below to predict
the decoherence dynamics exhibited by the numerical solution of the QLBE in
the limit $M \gg m$.

\subsubsection{Born approximation \label{sec:BA}}

Another simplification results when the interaction potential $V \left(
\tmmathbf{x} \right)$ is much weaker than the kinetic energy $E = p^2 / 2 m$.
One may then replace the exact scattering amplitude $f$ by its  Born
approximation $f_B^{}$, which is determined by the Fourier transform of the
interaction potential, 
\begin{eqnarray}
  f_B \left( \tmmathbf{p}_f -\tmmathbf{p}_i \right) & = & - \frac{m_{\ast}}{2
  \pi \hbar^2} \int \mathd \tmmathbf{x} \bignone V \left( \tmmathbf{x} \right)
  \exp \left( - i \frac{\left( \tmmathbf{p}_f -\tmmathbf{p}_i \right) \cdot
  \tmmathbf{x}}{\hbar} \right)  \label{eq:Bornf1} \, .
  \label{eq:Bornf2}
\end{eqnarray}
The approximated scattering amplitude therefore depends on the momentum transfer
$\tmmathbf{p}_f -\tmmathbf{p}_i$ only, so that the function $f$ in Eq.~(\ref{eq:QLBEL}) is not operator-valued anymore. Taking $\mu$ to be the Maxwell-Boltzmann distribution
\begin{eqnarray}
  \mu \left( {\tmmathbf{p}} \right) & = & \frac{1}{\left( 2 \pi mkT \right)^{3 /
  2}} \, \exp \left( - \frac{\left| {\tmmathbf{p}} \right|^2}{2 mkT} \right) \,,
  \label{eq:DCL2}
\end{eqnarray}
one may then perform the ${\tmmathbf{k}_{\bot}}$-integration in (\ref{eq:QLBE}), such
that the dissipator $\mathcal{D}$ defined by Eq.~(\ref{eq:QLBE}) becomes
{\cite{Vacchini2001b,Vacchini2002a,Hornberger2008a,Vacchini2009a}}
\begin{eqnarray}
  \mathcal{D}_B \rho & = & \int \mathd \tmmathbf{Q} \bignone \left(
  e^{i\tmmathbf{Q} \cdot \mathsf{X} / \hbar} L_B \left( \mathsf{P},
  \tmmathbf{Q} \right) \rho L_B^{\dag} \left( \mathsf{P}, \tmmathbf{Q} \right)
  e^{- i\tmmathbf{Q} \cdot \mathsf{X} / \hbar} \right. \nonumber\\
  &  & \left. - \frac{1}{2}  \left\{ \rho, L_B^{\dag} \left( \mathsf{P},
  \tmmathbf{Q} \right) L_B \left( \mathsf{P}, \tmmathbf{Q} \right) \right\}
  \bignone \right) \bignone \bignone \, .  \label{eq:BornD}
\end{eqnarray}
Here the Lindblad operators contain the functions $L_B \left( \mathsf{P},
\tmmathbf{Q} \right)$,  given by the expression
{\cite{Vacchini2001b,Vacchini2002a,Vacchini2009a}}
\begin{eqnarray}
  L_B \left( \mathbf{P}, \tmmathbf{Q} \right) & = & \left( \frac{\beta m}{2
  \pi} \right)^{1 / 4} \sqrt{\frac{n_{\tmop{gas}} \sigma_B \left( \tmmathbf{Q}
  \right)}{m_{\ast}^2 Q}} \nonumber\\
  &  & \times \exp \left( - \frac{\beta}{16 mQ^2} \left[ \left( 1 +
  \frac{m}{M} \right) Q^2 + 2 \frac{m}{M} \mathbf{P} \cdot \tmmathbf{Q}
  \right]^2 \right) \,,  \label{eq:BornL}
\end{eqnarray}
where $\sigma_B \left( \tmmathbf{Q} \right) \equiv \left| f_B \left(
\tmmathbf{Q} \right) \right|^2$ denotes the differential cross section in Born
approximation and $\beta \equiv 1 / kT$ is the inverse temperature.
The QLBE in Born approximation defined by Eqs.~(\ref{eq:BornD}) and
(\ref{eq:BornL}) was first proposed by Vacchini in Refs.~{\cite{Vacchini2001b,Vacchini2002a}}. As already mentioned, its solution may be obtained numerically by the stochastic simulation algorithm constructed in Ref.~{\cite{Breuer2007a}}.

\subsubsection{The limit of quantum Brownian motion \label{sec:DL}}

The {\tmem{quantum Brownian motion}} or {\tmem{diffusion limit}} applies when the state of the test particle is close to a thermal equilibrium state and when its mass is much greater than the mass of the gas particles {\cite{Hornberger2008a,Vacchini2009a}}. The momentum transfer $\tmmathbf{Q}$ is then small compared to the momentum of
the tracer particle. As discussed in {\cite{Vacchini2007a}}, this permits the expansion of the Lindblad operators in (\ref{eq:QLBE}) up to second order in the position and momentum operators.  
This expansion yields the
Caldeira-Leggett equation \cite{Caldeira1983a,Breuer2007b} in the minimally extended form as required to ensure a  Lindblad structure \cite{Vacchini2007a,Vacchini2009a},
\begin{eqnarray}
  \frac{\mathd}{\mathd t} \rho & = & \frac{1}{i \hbar} \left[ \mathsf{H}_S,
  \rho \right] + \frac{\gamma}{i \hbar} \left[ \mathsf{X}, \mathsf{P} \rho +
  \rho \mathsf{P} \right] - \frac{4 \pi \gamma}{\Lambda_{\tmop{th}}^2} \left[
  \mathsf{X}, \left[ \mathsf{X}, \rho \right] \right] \nonumber\\
  &  & - \frac{\gamma \Lambda_{\tmop{th}}^2}{16 \pi \hbar^2} \left[
  \mathsf{P}, \left[ \mathsf{P}, \rho \right] \right] \, .  \label{eq:CL}
\end{eqnarray}
Here, $\Lambda_{\tmop{th}}^2=2\pi\hbar^2\beta/M$ gives the thermal de Broglie wave length, and $\gamma$ is the relaxation rate. It is remarkable that the derivation leads to a microscopic expression for the latter {\cite{Vacchini2009a}},
\begin{eqnarray}
  \gamma & = & n_{\tmop{gas}} \frac{8 m}{3 M} \sqrt{\frac{2 \pi}{m \beta}}
  \int_0^{\infty} \mathd u \bignone u^5 e^{- u^2} \int_0^{\pi} \mathd \theta
  \sin \bignone \theta \left( 1 - \cos \theta \right) \left| f \left( \cos\theta,
  up_{\beta} \right) \right|^2 \, . \nonumber\\
  &  &  \label{eq:relaxationrate}
\end{eqnarray}
The velocities of the gas particles are here assumed to be Maxwell-Boltzmann
distributed, and the scattering to be isotropic so that the amplitude $f$ depends only on the
scattering angle $\theta$ and the modulus of the momentum $p \equiv \left|
\tmmathbf{p}_i \right| = \left| \tmmathbf{p}_f \right|$. The integration variable $u \equiv p / p_{\beta}$ denotes the momentum in dimensionless form, where $p_{\beta} = \sqrt{2 m / \beta}$ is the most probable momentum at temperature $T = 1 / \left( k_B \beta \right)$.

\section{Monte Carlo unraveling \label{sec:MCU}}

To solve the QLBE we now employ the Monte Carlo wave function method
{\cite{Molmer1993a,Molmer1996a,Gardiner1992a,Carmichael1993a,Breuer2007b}}. The underlying idea of this approach is to regard the wave function as a stochastic process in the Hilbert space of pure system states, with the property that the expectation value
$
  \rho \left( t \right)  =  \mathbb{E} \left[ | \psi \left( t \right)
  \rangle \langle \psi \left( t \right) |  \label{eq:MC1} \right]
$
satisfies a given Lindblad master equation, 
$\partial_t\rho=(i\hbar)^{-1}[\mathsf{H},\rho]+\sum_i(\mathsf{L}_i\rho\mathsf{L}_i^\dagger-\frac{1}{2}\{\mathsf{L}_i^\dagger\mathsf{L}_i,\rho\})$. Any process with this property is called an unraveling of the master equation. An appropriate stochastic differential equation defining such a process is given by \cite{Breuer2007b}
\begin{eqnarray}
  |d \psi_t \rangle & = & - \frac{i}{\hbar} \mathsf{H}_{\tmop{eff}} | \psi_t
  \rangle \mathd t + \bignone \frac{1}{2} \sum_i \gamma_i \bignone \bignone \|
  \mathsf{L}_i | \psi_t \rangle \|^2 | \psi_t \rangle \mathd t \nonumber\\
  &  & + \sum_i \left( \frac{\mathsf{L}_i | \psi_t \rangle}{\| \mathsf{L}_i |
  \psi_t \rangle \|} - | \psi_t \rangle \right) \mathd N_i \left( t \right)
  \,,  \label{eq:MCSDE}
\end{eqnarray}
where $\mathsf{H}_{\tmop{eff}}$ represents the non-Hermitian
operator
\begin{eqnarray}
  \mathsf{H}_{\tmop{eff}} & = & \mathsf{H} - \frac{i \hbar}{2} \sum_i \gamma_i
  \mathsf{L}_i^{\dag} \mathsf{L}_i \bignone \, .  \label{eq:MCHeff}
\end{eqnarray}
The random Poisson increments $\mathd N_i \left( t \right)$ in Eq.~(\ref{eq:MCSDE}) satisfy the relations
\begin{eqnarray}
  \mathd N_i \left( t \right) \mathd N_j \left( t \right) & = & \delta_{ij}
  \mathd N_i \left( t \right) \,,  \label{eq:ITOPDP}
\end{eqnarray}
and their expectation values are given by
\begin{eqnarray}
  \mathbb{E} \left[ \mathd N_i \left( t \right) \right] & = & \gamma_i \|
  \mathsf{L}_i | \psi_t \rangle \|^2 \mathd t \bignone \, .
  \label{eq:MCrates}
\end{eqnarray}
The Monte Carlo method consists of generating an ensemble of realizations
$\left\{ | \psi_{\alpha} \left( t \right) \rangle \right\}$ of the process defined by the stochastic differential equation (\ref{eq:MCSDE}), and of estimating the density matrix $\rho \left( t
\right)$ through an ensemble average {\cite{Molmer1993a,Molmer1996a,Gardiner1992a,Carmichael1993a,Breuer2007b}}.

In the following, we  briefly summarize a general algorithm which is often used for the numerical implementation of the stochastic differential equation (\ref{eq:MCSDE}) {\cite{Gardiner1992a}}. 
This method forms the basis for the stochastic algorithm presented below, which extends the procedures presented in Ref.~\cite{Breuer2007a}.

\subsection{The general algorithm \label{sec:AIG}}

We start with the normalized state $| \psi_t \rangle$ which has been reached through a quantum jump at time $t$ (or is the initial state). Subsequently, the state follows a deterministic time evolution which is given by the nonlinear equation of motion
\begin{eqnarray}
  \partial_t | \psi_t \rangle & = & - \frac{i}{\hbar} \mathsf{H}_{\tmop{eff}}
  | \psi_t \rangle + \bignone \frac{1}{2} \sum_i \gamma_i \bignone \bignone \|
  \mathsf{L}_i | \psi_t \rangle \|^2 | \psi_t \rangle  \label{eq:MCDet0}
\end{eqnarray}
with the formal solution
\begin{eqnarray}
  | \psi_{t + \tau} \rangle & = & \frac{\exp \left( - i
  \mathsf{H}_{\tmop{eff}} \tau / \hbar \right) | \psi_t \rangle}{\| \exp
  \left( - i \mathsf{H}_{\tmop{eff}} \tau / \hbar \right) | \psi_t \rangle \|} \, . \label{eq:MCDet1}
\end{eqnarray}
The probability for a jump to occur out of this state is
characterized by the total jump rate
\begin{eqnarray}
&&  \Gamma \left( \psi_t \right)  =  \frac{1}{\mathd t \bignone \bignone}
  \sum_i \mathbb{E} \left[ \mathd N_i \left( t \right) \right] 
   =  \sum_i \gamma_i \bignone \, \| \mathsf{L}_i | \psi_t \rangle \|^2.
  \label{eq:MCGamma}
\end{eqnarray}
It allows one to evaluate the corresponding {\tmem{waiting time distribution}}
$W \left( \tau | \psi_t \right)$, the cumulative distribution function representing the probability that a jump occurs in the time interval $\left[t,t+\tau\right]$, 
\begin{eqnarray}
  W \left( \tau | \psi_t \right) & = & 1 -\| \exp \left( - i
  \mathsf{H}_{\tmop{eff}} \tau / \hbar \right) | \psi_t \rangle \|^2 \, .
  \label{eq:MCWait}
\end{eqnarray}
In practice, a realization $\tau$ of the random waiting time can be obtained by the inversion method, i.e. by numerically solving the equation
\begin{eqnarray}
  \eta & = & \| \exp \left( - i \mathsf{H}_{\tmop{eff}} \tau / \hbar \right) |
  \psi_t \rangle \|^2  \label{eq:MCtau}
\end{eqnarray}
for $\tau$, with $\eta$  a random number drawn uniformly from the interval $\left[0, 1 \right]$.
At time $t+\tau$ a discontinuous quantum jump occurs, i.e. the wave function  $|\psi_{t+\tau}\rangle$ is replaced according to
\begin{eqnarray}
  | \psi_{t + \tau} \rangle & \rightarrow & \frac{\mathsf{L}_i | \psi_{t +
  \tau} \rangle}{\| \mathsf{L}_i | \psi_{t + \tau} \rangle \|} \, .
  \label{eq:MCjumps}
\end{eqnarray}
The corresponding jump operator, labeled by the index $i$, is drawn from the probability
distribution given by the ratio of the jump rate $\Gamma_i \left( \psi_{t +
\tau} \right) =\mathbb{E} \left[ \mathd N_i \left( t + \tau \right) \right] /
\mathd t \bignone$ of the Poisson process $N_i \left( t \right)$ and the total
jump rate $\Gamma \left( \psi_{t + \tau} \right)$, 
\begin{eqnarray}
  &&\tmop{Prob} \left( i| \psi_{t + \tau} \right) =  \frac{\Gamma_i \left(
  \psi_{t + \tau} \right)}{\Gamma \left( \psi_{t + \tau} \right)} 
   =  \frac{\gamma_i \|
  \mathsf{L}_i | \psi_{t + \tau} \rangle \|^2}{\Gamma \left( \psi_{t + \tau} \right)}  \: .  \label{eq:MCprob}
\end{eqnarray}

\subsection{Unraveling the QLBE \label{sec:AQLBE}}

We now adapt the Monte Carlo method to solve the QLBE, which is
characterized by the family of Lindblad operators
$e^{i\tmmathbf{Q} \cdot \mathsf{X} / \hbar} L \left( {\tmmathbf{k}_{\bot}}, \mathsf{P},
\tmmathbf{Q} \right)$. For this purpose, the index $i$ is replaced by the continuous variables $\tmmathbf{Q}$ and ${\tmmathbf{k}_{\bot}}$,
and the sums over $i$ are substituted by the integrals
\begin{eqnarray}
  \sum_i \bignone & \rightarrow & \int_{\mathbb{R}^3} \mathd \tmmathbf{Q}
  \int_{\tmmathbf{Q}^{\perp}} \mathd {\tmmathbf{k}_{\bot}} \, .  \label{eq:repl2}
\end{eqnarray}
Although the procedure is straightforward, we repeat here the main steps since the obtained formulas are required for reference later on.

The Monte Carlo unraveling of the QLBE is described by the stochastic Schr\"odinger equation
\begin{eqnarray}
  |d \psi_t \rangle & = & - \frac{i}{\hbar} \mathsf{H}_{\tmop{eff}} | \psi_t
  \rangle \mathd t + \bignone \frac{1}{2} \int_{\mathbb{R}^3} \mathd
  \tmmathbf{Q} \int_{\tmmathbf{Q}^{\perp}} \mathd {\tmmathbf{k}_{\bot}} \bignone
  \bignone \bignone \bignone \|L \left( {\tmmathbf{k}_{\bot}}, \mathsf{P}, \tmmathbf{Q}
  \right) | \psi_t \rangle \|^2 | \psi_t \rangle \mathd t
  \label{eq:QLBESDE}\\
  &  & + \int_{\mathbb{R}^3} \mathd \tmmathbf{Q} \int_{\tmmathbf{Q}^{\perp}}
  \mathd {\tmmathbf{k}_{\bot}} \left( \frac{e^{i\tmmathbf{Q} \cdot \mathsf{X} / \hbar}
  L \left( {\tmmathbf{k}_{\bot}}, \mathsf{P}, \tmmathbf{Q} \right) | \psi_t
  \rangle}{\|L \left( {\tmmathbf{k}_{\bot}}, \mathsf{P}, \tmmathbf{Q} \right) | \psi_t
  \rangle \|} - | \psi_t \rangle \right) \mathd N_{\tmmathbf{Q}, {\tmmathbf{k}_{\bot}}}
  \left( t \right) \,, \nonumber
\end{eqnarray}
where the effective Hamiltonian has the form
\begin{eqnarray}
  \mathsf{H}_{\tmop{eff}} & = & \mathsf{H} - \frac{i \hbar}{2}
  \int_{\mathbb{R}^3} \mathd \tmmathbf{Q} \int_{\tmmathbf{Q}^{\perp}} \mathd
  {\tmmathbf{k}_{\bot}} \bignone \mathsf{L}^{\dag} \left( {\tmmathbf{k}_{\bot}}, \mathsf{P},
  \tmmathbf{Q} \right) \mathsf{L} \left( {\tmmathbf{k}_{\bot}}, \mathsf{P},
  \tmmathbf{Q} \right) \bignone \, .  \label{eq:QLBHeff}
\end{eqnarray}
The Poisson increments in (\ref{eq:QLBESDE}) have the expectation values
\begin{eqnarray}
  \mathbb{E} \left[ \mathd N_{\tmmathbf{Q}, {\tmmathbf{k}_{\bot}}} \left( t \right)
  \right] & = & \| \mathsf{L} \left( {\tmmathbf{k}_{\bot}}, \mathsf{P}, \tmmathbf{Q}
  \right) | \psi_t \rangle \|^2 \mathd t \,,  \label{eq:QLBExp}
\end{eqnarray}
and satisfy
\begin{eqnarray}
  \mathd N_{\tmmathbf{Q}, {\tmmathbf{k}_{\bot}}} \left( t \right) \mathd
  N_{\tmmathbf{Q}', {\tmmathbf{k}_{\bot}'}} \left( t \right) & = & \delta^{\left( 3
  \right)} \left( \tmmathbf{Q}-\tmmathbf{Q}' \right) \delta^{\left( 2 \right)}
  \left( {\tmmathbf{k}_{\bot}} -{\tmmathbf{k}_{\bot}'}
  \right) \mathd N_{\tmmathbf{Q}, {\tmmathbf{k}_{\bot}}} \left( t \right) \, .
  \label{eq:QLBIto}
\end{eqnarray}
These relations represent the continuous counterpart of the discrete set of equations (\ref{eq:ITOPDP}).
The deterministic part of the Monte Carlo unraveling is generated by the
nonlinear equation
\begin{eqnarray}
  \partial_t | \psi_t \rangle & = & - \frac{i}{\hbar} \mathsf{H}_{\tmop{eff}}
  | \psi_t \rangle + \bignone \frac{1}{2} \int_{\mathbb{R}^3} \mathd
  \tmmathbf{Q} \int_{\tmmathbf{Q}^{\perp}} \mathd {\tmmathbf{k}_{\bot}} \bignone
  \bignone \bignone \bignone \|L \left( {\tmmathbf{k}_{\bot}}, \mathsf{P}, \tmmathbf{Q}
  \right) | \psi_t \rangle \|^2 | \psi_t \rangle \,,  \label{eq:QLBDiff}
\end{eqnarray}
whose formal solution is given by Eq.~(\ref{eq:MCDet1}). The 
jump probability is determined by the rate
\begin{eqnarray}
  \Gamma \left( \psi_t \right) & = & \int_{\mathbb{R}^3} \mathd \tmmathbf{Q}
  \int_{\tmmathbf{Q}^{\perp}} \mathd {\tmmathbf{k}_{\bot}}\|L \left( {\tmmathbf{k}_{\bot}},
  \mathsf{P}, \tmmathbf{Q} \right) | \psi_t \rangle \|^2 \,,
  \label{eq:QLBGamma}
\end{eqnarray}
and a realization of the random waiting time $\tau$ is obtained by  solving Eq.~(\ref{eq:MCtau}) for $\tau$ with the effective Hamiltonian (\ref{eq:QLBHeff}). The jump at time $t + \tau$ is effected by
\begin{eqnarray}
  | \psi \left( t + \tau \right) \rangle & \rightarrow &
  \frac{e^{i\tmmathbf{Q} \cdot \mathsf{X} / \hbar} L \left( {\tmmathbf{k}_{\bot}},
  \mathsf{P}, \tmmathbf{Q} \right) | \psi_{t + \tau} \rangle}{\|L \left(
  {\tmmathbf{k}_{\bot}}, \mathsf{P}, \tmmathbf{Q} \right) | \psi_{t + \tau} \rangle \|}
  \, ,  \label{eq:QLBjumps}
\end{eqnarray}
where the continuous parameters ${\tmmathbf{k}_{\bot}}$ and $\tmmathbf{Q}$ characterizing the jump
operator are drawn from the probability density
\begin{eqnarray}
  \tmop{Prob} \left( {\tmmathbf{k}_{\bot}}, \tmmathbf{Q}| \psi_{t + \tau} \right) & = &
  \frac{1}{\Gamma \left( | \psi_{t + \tau} \rangle \right)} \|L \left(
  {\tmmathbf{k}_{\bot}}, \mathsf{P}, \tmmathbf{Q} \right) | \psi_{t + \tau} \rangle
  \|^2 \, .  \label{eq:QLBProb}
\end{eqnarray}

\subsection{Unraveling the QLBE in the momentum basis \label{sec:UQLBEM}}

 The implementation of the above
algorithm is particularly simple when the initial state is a discrete
superposition of a finite number of momentum eigenstates {\cite{Breuer2007a}},
\begin{eqnarray}
  | \psi \left( 0 \right) \rangle & = & \sum_{i = 1}^N \alpha_i \left( 0
  \right) |\tmmathbf{P}_i \left( 0 \right) \rangle \,, \hspace{1em}
  \tmop{with} \: \bignone \sum_{i = 1}^N \left| \alpha_i \left( 0 \right)
  \right|^2 = 1 \, .  \label{eq:MB0}
\end{eqnarray}
Due to the translation-covariance of the QLBE the Lindblad operators have the
structure $e^{i\tmmathbf{Q} \cdot \mathsf{X} / \hbar} L \left( {\tmmathbf{k}_{\bot}},
\mathsf{P}, \tmmathbf{Q} \right)$. This implies that the effective Hamiltonian
is a function of the momentum operator only, so that the deterministic
evolution of (\ref{eq:MB0}) affects solely the amplitudes of the superposition, that is
\begin{eqnarray}
  &  & \text{$| \psi \left( t \right) \rangle = \sum_{i = 1}^N \alpha_i
  \left( t \right) |\tmmathbf{P}_i \left( 0 \right) \rangle$} \, .
  \label{eq:MB1}
\end{eqnarray}
The jumps, on the other hand, cause a translation of the momentum eigenstates
and a redistribution of the amplitudes,
\begin{eqnarray}
 e^{i\tmmathbf{Q} \cdot \mathsf{X} / \hbar} L \left(
  {\tmmathbf{k}_{\bot}}, \mathsf{P}, \tmmathbf{Q} \right) | \psi \left( t \right)
  \rangle & = &  \sum_{i = 1}^N \alpha'_i \left( t \right)
  |\tmmathbf{P}_i +\tmmathbf{Q} \rangle \,.  \label{eq:MB2}
\end{eqnarray}
This shows that the quantum
trajectory $| \psi(t) \rangle$ remains a superposition of $N$ momentum eigenstates at all times. The stochastic process therefore reduces to a process in the
finite-dimensional space of the amplitudes $\alpha_i$ and
momenta $\tmmathbf{P}_i$. Here the momentum eigenstates  are
taken to be normalized with respect to a large volume $\Omega$, such that they
form a discrete basis,  $\langle \tmmathbf{P}_i |\tmmathbf{P}_j \rangle =
\delta_{ij}$. 

In the following it is convenient to work with
dimensionless variables
\begin{eqnarray}
  \tmmathbf{U} \equiv \frac{\tmmathbf{P}}{Mv_{\beta}} \,, \hspace{1em}
  \tmmathbf{K} \equiv \frac{\tmmathbf{Q}}{m_{\ast} v_{\beta}} \,, \hspace{1em}
  \tmmathbf{W}_{\bot} \equiv \frac{{\tmmathbf{k}_{\bot}}}{mv_{\beta}} \,, &  &
  \label{eq:QLBKW}
\end{eqnarray}
where the scale is given by the most probable velocity of the gas particles
$v_{\beta} = \sqrt{2 k_B T / m}$. Note that $\tmmathbf{W}_{\bot}$, being proportional
to ${\tmmathbf{k}_{\bot}}$, lies in the plane perpendicular to
$\tmmathbf{K}$. The quantum trajectories are then represented as
\begin{eqnarray}
  | \psi \left( t \right) \rangle & = & \sum_{i = 1}^N \alpha_i \left( t
  \right) |\tmmathbf{U}_i \left( t \right) \rangle \,, \hspace{1em}
  \tmop{with} \: \bignone \sum_{i = 1}^N \left| \alpha_i \left( t \right)
  \right|^2 = 1 \, .  \label{eq:QLBinitial}
\end{eqnarray}

Before discussing the unraveling of the QLBE in more
detail, let us evaluate the jump rate (\ref{eq:QLBGamma}) for momentum
eigenstates, $| \psi_t \rangle = |\tmmathbf{P} \rangle$. This quantity appears frequently in the algorithm described below.
By inserting $| \psi_t \rangle = |\tmmathbf{P} \rangle$ into
Eq.~(\ref{eq:QLBGamma}), one obtains
\begin{eqnarray}
  \Gamma \left( \tmmathbf{P} \right) & = &  \int_{\mathbb{R}^3} \mathd \tmmathbf{Q} \int_{\tmmathbf{Q}^{\perp}}
  \mathd {\tmmathbf{k}_{\bot}}|L \left( {\tmmathbf{k}_{\bot}}, \tmmathbf{P}, \tmmathbf{Q}
  \right) |^2 \, . \,  \label{eq:GammaP}
\end{eqnarray}
Noting Eq.~(\ref{eq:W1}), one finds that the jump rate agrees with the total
collision rate for a particle with momentum $\tmmathbf{P}$,
\begin{eqnarray}
  \Gamma \left( \tmmathbf{P} \right)_{} & = & \int \mathd \tmmathbf{Q}
  \bignone M^{\tmop{cl}} \left( \tmmathbf{P} \rightarrow
  \tmmathbf{P}+\tmmathbf{Q} \right) \, .  \label{eq:totalcollrate1}
\end{eqnarray}
It follows that $\Gamma \left( \tmmathbf{P} \right) = \Gamma \left( P \right)$
is a function of the modulus of $\tmmathbf{P}$ only, since the collision rate
must be independent of the orientation of $\tmmathbf{P}$ for a homogeneous
background gas.
Upon using the dimensionless quantities (\ref{eq:QLBKW}), and after inserting (\ref{eq:QLBEL}) for $L$, as well as the Maxwell-Boltzmann distribution
(\ref{eq:DCL2}), one finds
\begin{eqnarray}
  \Gamma \left( U \right) & = & \int_{\mathbb{R}^3} \mathd \tmmathbf{K}
  \int_{\tmmathbf{K}^{\perp}} \mathd \tmmathbf{W}_{\bot}g \left( \tmmathbf{W}_{\bot},
  \tmmathbf{U}, \tmmathbf{K} \right) p_{\sigma_K} \left( \tmmathbf{K} \right)
  p_{\sigma_W} \left( \tmmathbf{W}_{\bot} \right) \,,  \label{eq:QLBG3}
\end{eqnarray}
with
\begin{eqnarray}
  g \left( \tmmathbf{W}_{\bot}, \tmmathbf{U}, \tmmathbf{K} \right) & = &  \frac{8 \pi
  n_{\tmop{gas}} v_{\beta}}{\left| \tmmathbf{K} \right|} \left| f
  \left( m_{\ast} v_{\beta} \left[ \tmmathbf{W}_{\bot} -\tmmathbf{U}_{\perp \tmmathbf{K}} - \frac{\tmmathbf{K}}{2}
  \right], m_{\ast} v_{\beta} \left[ \tmmathbf{W}_{\bot} -\tmmathbf{U}_{\perp \tmmathbf{K}} + \frac{\tmmathbf{K}}{2}
  \right] \right) \right|^2 \nonumber\\
  &  & \times e^{-\tmmathbf{K} \cdot \tmmathbf{U}}
  e^{-\tmmathbf{U}^2_{\|\tmmathbf{K}}} \,.  \label{eq:QLBG4}
\end{eqnarray}
The densities $p_{\sigma_K} \left(
\tmmathbf{K} \right)$ and $p_{\sigma_W} \left( \tmmathbf{W}_{\bot} \right)$ denote
three- and two-dimensional normal distributions, respectively,
\begin{eqnarray}
  p_{\sigma_K} \left( \tmmathbf{K} \right) & = & \frac{1}{\left( 2 \pi
  \sigma_K \right)^{3 / 2}} \exp \left( - \frac{\tmmathbf{K}^2}{2 \sigma_K^2}
  \right) \,, \nonumber\\
  p_{\sigma_W} \left( \tmmathbf{W}_{\bot} \right) & = & \frac{1}{2 \pi \sigma_W} \exp
  \left( - \frac{\tmmathbf{W}_{\bot}^2}{2 \sigma_W^2} \right) \,,  \label{eq:QLBG5}
\end{eqnarray}
with variances $\sigma_K = \sqrt{2}$ and $\sigma_W = 1 / \sqrt{2}$.

The integral (\ref{eq:QLBG3}) can be evaluated numerically using a Monte Carlo
method with importance sampling {\cite{Press2007a}}. For this purpose, one
draws $n$ samples $\tmmathbf{K}_i$ from the normal distribution $p_{\sigma_K}
\left( \tmmathbf{K} \right)$ and computes orthonormal vectors $\tmmathbf{e}_{1
i}$ and $\tmmathbf{e}_{2 i}$ which are orthogonal to $\tmmathbf{K}_i$, i.e.
$
   \tmmathbf{e}_{1 i} \cdot \tmmathbf{K}_i = 0$,
  $\tmmathbf{e}_{2 i} \cdot \tmmathbf{K}_i = 0$, $\tmmathbf{e}_{1 i} \cdot
  \tmmathbf{e}_{2 i} = 0 $
using the Gram-Schmidt method. As a next step, $n$ further samples $\left(
u_i, v_i \right)$ are drawn from the two-dimensional Gaussian distribution
$p_{\sigma_W}$, which yields a sample of scaled momentum vectors
$
  \tmmathbf{W}_{i \bot}  =  u_i \tmmathbf{e}_{1 i} + v_i  \tmmathbf{e}_{2 i}.
$
The jump rate (\ref{eq:QLBG3}) is then approximated by the average
\begin{eqnarray}
  \Gamma \left( U \right) & \simeq & \frac{1}{n} \sum^n_{i = 1} g \left(
  \tmmathbf{W}_{i \bot}, \tmmathbf{U}, \tmmathbf{K}_i \right) \, .
  \label{eq:Importancesample}
\end{eqnarray}

Let us now discuss in more detail the unraveling of the QLBE in the momentum
basis. To this end, suppose the state
\begin{eqnarray}
  | \psi \left( t \right) \rangle & = & \sum_{i = 1}^N \alpha_i \left( t
  \right) |\tmmathbf{U}_i \left( t \right) \rangle   \label{eq:QLBDet0}
\end{eqnarray}
was obtained through a quantum jump at time $t$. As mentioned above, the
effective Hamiltonian (\ref{eq:QLBHeff}) depends on the momentum operator
only, so that the momenta $\tmmathbf{U}_i$ stay constant during the
deterministic evolution. The propagation of the state (\ref{eq:QLBDet0}) with
the non-Hermitian operator (\ref{eq:QLBHeff}) thus yields \cite{Breuer2007a}
\begin{eqnarray}
  | \psi \left( t + \tau \right) \rangle & = & \sum_{i = 1}^N \alpha_i \left(
  t + \tau \right) |\tmmathbf{U}_i \left( t \right) \rangle \, .
  \label{eq:QLBDet1}
\end{eqnarray}
Here the weights have the form
\begin{eqnarray}
  \alpha_i \left( t + \tau \right) & = & \frac{1}{\mathcal{N}} \, \exp \left(
  - \frac{i}{2 \hbar} Mv_{\beta} \tmmathbf{U}_i^2 \tau \right) \exp \left( -
  \frac{\tau}{2} \Gamma \left( U_i \right) \right) \alpha_i \left( t \right)
  \,,  \label{eq:QBLDet2}
\end{eqnarray}
with the normalization
\begin{eqnarray}
  &  & \mathcal{N}^2  \: = \: \sum_{i = 1}^N \left| \alpha_i \left( t + \tau
  \right) \right|^2  \: = \: \sum_{i = 1}^N \left| \alpha_i \left( t \right)
  \right|^2 \exp \left( - \tau \Gamma \left( U_i \right) \right) \, .
  \label{eq:QLBDet3}
\end{eqnarray}

As a next step, one must evaluate the waiting times $\tau$. For this purpose,
consider the expression
\begin{eqnarray}
 \text{$\| \exp \left( - i \mathsf{H}_{\tmop{eff}} \tau / \hbar \right)
  | \psi_t \rangle \|^2$} & =& \sum_{i, j = 1}^N \alpha_i^{\ast} \left( t \right) \alpha_j
  \left( t \right) \langle \tmmathbf{U}_i \left( t \right) |e^{i
  \mathsf{H}^{\dag}_{\tmop{eff}} \tau / \hbar} e^{- i \mathsf{H}_{\tmop{eff}}
  \tau / \hbar} |\tmmathbf{U}_j \left( t \right) \rangle \, .
  \label{eq:QLBWait1}
\end{eqnarray}
By using the definition of $\mathsf{H}_{\tmop{eff}}$ (\ref{eq:QLBHeff}), the
fact that the two summands in $\mathsf{H}_{\tmop{eff}}$ commute, and the jump
rate (\ref{eq:QLBG3}), this yields
\begin{eqnarray}
  \text{$\| \exp \left( - i \mathsf{H}_{\tmop{eff}} \tau / \hbar \right) |
  \psi_t \rangle \|^2$} & = & \sum_{i = 1}^N \left| \alpha_i \left( t \right)
  \right|^2 \exp \left( - \tau \Gamma \left( U_i \right) \right) \, .
  \label{eq:QLBWait2}
\end{eqnarray}
It follows from (\ref{eq:MCtau}) that samples of the waiting times $\tau$ are
obtained by numerically inverting the non-algebraic equation
\begin{eqnarray}
  \eta & = & \sum_{i = 1}^N \left| \alpha_i \left( t \right) \right|^2 \exp
  \left( - \tau \Gamma \left( U_i \right) \right) \,,  \label{eq:QLBWait3}
\end{eqnarray}
with $\eta$ drawn from the uniform distribution on $\left[ 0, 1 \right]$.

To be able to carry out the quantum jumps, we have to determine the momentum parameters
$\tmmathbf{K}$ and $\tmmathbf{W}_{\bot}$, which characterize the jump operator. These
vectors are obtained by sampling from the probability distribution
(\ref{eq:QLBProb}). Upon inserting states of the form (\ref{eq:QLBDet1}),
Eq.~(\ref{eq:QLBProb}) becomes
\begin{eqnarray}
  \tmop{Prob} \left( \tmmathbf{W}_{\bot}, \tmmathbf{K}| \psi_{t + \tau} \right) & = &
  \sum_{i = 1}^N \frac{\left| \alpha_i \left( t + \tau \right) \right|^2
  \Gamma \left( U_i \right)}{\sum_{j = 1}^N \bignone \left| \alpha_j \left( t
  + \tau \right) \right|^2 \Gamma \left( U_j \right)} \frac{\|L \left(
  \tmmathbf{W}_{\bot}, \tmmathbf{U}_i, \tmmathbf{K} \right) |\tmmathbf{U}_i \rangle
  \|^2}{\Gamma \left( U_i \right)} \nonumber\\
  & \equiv & \sum_{i = 1}^N p_i \, \tmop{Prob} \left( \tmmathbf{W}_{\bot},
  \tmmathbf{K}|\tmmathbf{U}_i \right)_{} \, .  \label{eq:Distr2}
\end{eqnarray}
This distribution is a mixture of the probabilities
\begin{eqnarray}
  p_i & = & \frac{\left| \alpha_i \left( t + \tau \right) \right|^2 \Gamma
  \left( U_i \right)}{\sum_{j = 1}^N \bignone \left| \alpha_j \left( t + \tau
  \right) \right|^2 \Gamma \left( U_j \right)} \,,  \label{eq:Distr3}
\end{eqnarray}
and the probability densities
\begin{eqnarray}
  \tmop{Prob} \left( \tmmathbf{W}_{\bot}, \tmmathbf{K}|\tmmathbf{U}_i \right)_{} \, &
  = & \frac{\|L \left( \tmmathbf{W}_{\bot}, \tmmathbf{U}_i, \tmmathbf{K} \right)
  |\tmmathbf{U}_i \rangle \|^2}{\Gamma \left( U_i \right)} \nonumber\\
  & = & \frac{8 \pi n_{\tmop{gas}} v_{\beta}}{\Gamma \left( U_i \right)
  \left| \tmmathbf{K} \right|} \left| f \left( m_{\ast} v_{\beta} \left[ \tmmathbf{W}_{\bot} -\tmmathbf{U}_{i\perp \tmmathbf{K}}-
  \frac{\tmmathbf{K}}{2} \right], m_{\ast} v_{\beta} \left[ \tmmathbf{W}_{\bot} -\tmmathbf{U}_{i\perp \tmmathbf{K}} +
  \frac{\tmmathbf{K}}{2} \right] \right) \right|^2 \nonumber\\
  &  & \times \mu \left( mv_{\beta} \left[ \tmmathbf{W}_{\bot}
   +\tmmathbf{U}_{i\|\tmmathbf{K}} + \frac{\tmmathbf{K}}{2}\right] \right) \, .
  \label{eq:Distr4}
\end{eqnarray}
In order to draw a sample from the mixture (\ref{eq:Distr3}), one may proceed
as follows {\cite{Breuer2007a}}. First, an index $i$ is drawn from the
probabilities (\ref{eq:Distr3}). Then, the momenta $\tmmathbf{K}$ and
$\tmmathbf{W}_{\bot}$ are drawn from the probability distribution $\tmop{Prob} \left(
\tmmathbf{W}_{\bot}, \tmmathbf{K}|\tmmathbf{U}_i \right)$ using a stochastic sampling
method, such as the {\tmem{Metropolis-Hastings algorithm}}
{\cite{Press2007a}}.

Having the momenta $\tmmathbf{W}_{\bot}$ and $\tmmathbf{K}$ at hand, one can now
perform the quantum jump. According to Eq.~(\ref{eq:QLBjumps}), the state
(\ref{eq:QLBDet1}) is transformed as
\begin{eqnarray}
  | \psi \left( t + \tau \right) \rangle & \rightarrow &
  \tilde{\mathcal{N}}^{- 1} \exp \left( \frac{i}{\hbar} mv_{\beta}
  \tmmathbf{K} \cdot \mathsf{X} \right) L \left( \tmmathbf{W}_{\bot}, \mathsf{U},
  \tmmathbf{K} \right) \sum_{i = 1}^N \alpha_i \left( t + \tau \right)
  |\tmmathbf{U}_i \left( t \right) \rangle \nonumber\\
  &  & = \, \sum_{i = 1}^N \tilde{\mathcal{N}}^{- 1} L \left( \tmmathbf{W}_{\bot},
  \tmmathbf{U}_i, \tmmathbf{K} \right) \alpha_i \left( t + \tau \right)
  |\tmmathbf{U}_i \left( t \right) + \frac{m_{\ast}}{M} \tmmathbf{K} \rangle
  \,,  \label{eq:jump0}
\end{eqnarray}
where the normalization $\tilde{\mathcal{N}}$ is determined by
\begin{eqnarray}
  \text{$\tilde{\mathcal{N}}$}^2 & = & \sum_{i = 1}^N \left| L \left(
  \tmmathbf{W}_{\bot}, \tmmathbf{U}_i, \tmmathbf{K} \right) \alpha_i \left( t + \tau
  \right) \right|^2 \bignone \, .  \label{eq:jumpnorm}
\end{eqnarray}
This shows that the momentum eigenstates are shifted
\begin{eqnarray}
  |\tmmathbf{U}_i \rangle & \rightarrow & |\tmmathbf{U}_i + \frac{m_{\ast}}{M}
  \tmmathbf{K} \rangle \,,  \label{eq:jumpmom}
\end{eqnarray}
while the weights are redistributed as
\begin{eqnarray}
  \alpha_i \left( t + \tau \right) & \rightarrow & \alpha_i' \left( t + \tau
  \right) = x_i \alpha_i \left( t + \tau \right) \,,  \label{eq:jumpw1}
\end{eqnarray}
where the factors $x_i$ are given by $x_i = \tilde{\mathcal{N}}^{- 1} L \left(
\tmmathbf{W}_{\bot}, \tmmathbf{U}_i, \tmmathbf{K} \right)$. Upon using the explicit
form (\ref{eq:QLBEL}) of $L$, and by inserting the Maxwell-Boltzmann
distribution (\ref{eq:DCL2}), we find
\begin{eqnarray}
  x_i & = & \frac{1}{\tilde{\mathcal{N}}} f \left( m_{\ast} v_{\beta} \left[
  \tmmathbf{W}_{\bot} -\tmmathbf{U}_{i \perp \tmmathbf{K}} - \frac{\tmmathbf{K}}{2} \right], m_{\ast} v_{\beta} \left[
  \tmmathbf{W}_{\bot} -\tmmathbf{U}_{i \perp \tmmathbf{K}} + \frac{\tmmathbf{K}}{2} \right] \right) \nonumber\\
  &  & \times \exp \left( - \frac{1}{2} \left[ \frac{\tmmathbf{K}}{2}
  +\tmmathbf{U}_{i\|\tmmathbf{K}} \right]^2 \right) \,.  \label{eq:jump3a}
\end{eqnarray}

According to Eq.~(\ref{eq:jumpmom}), the momentum eigenstates are all shifted
with the same momentum $\tmmathbf{K}$ in a quantum jump. This fact is decisive for the numerical performance of the algorithm, since it implies
that the time consuming Metropolis-Hastings algorithm must be applied only
once for all $i \in \left\{ 1, \ldots, N \right\}$. This suggests that the
algorithm can be applied also to initial states which are superpositions of
many momentum eigenstates.

\begin{figure}[tb]
  \resizebox{9cm}{!}{\includegraphics{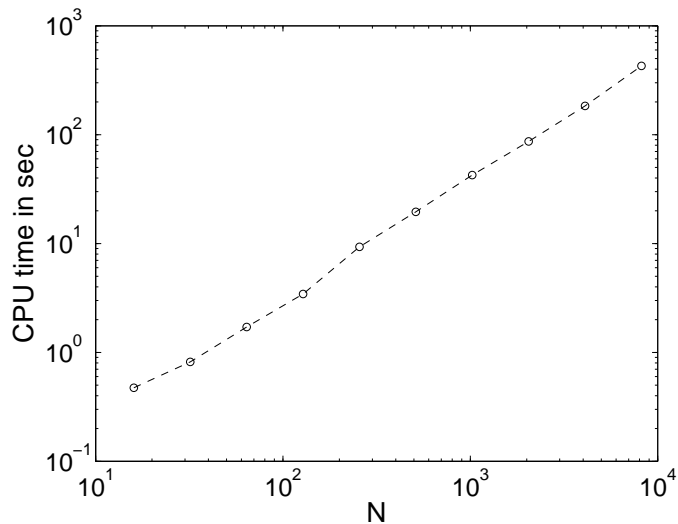}}
  \caption{CPU time of the Monte Carlo unraveling as a function of the number
  $N$ of basis states involved in the initial superposition. The curve is
  almost a straight line with slope $a \simeq 1.1$ in the logarithmic plot. It
  follows that the CPU time grows almost linearly with $N$, that is $T \propto
  N^{1.1}$. \label{fig:CPU}}
\end{figure}

This fact is substantiated by the numerical analysis depicted in the logarithmic
plot of Fig.~\ref{fig:CPU}. Here, the CPU time of the above algorithm is shown as a function of the number $N$ of basis states involved in the
initial superposition state $| \psi
\left( 0 \right) \rangle = \sum_{i = 1}^N |\tmmathbf{U}_i \rangle / \sqrt{N}
\bignone$, with $\tmmathbf{U}_i = \left( 0, 0, i \right)$. The simulation is
based on $10^2$ quantum trajectories in each run. The curve shown in
Fig.~\ref{fig:CPU} is almost a straight line with a slope $a \simeq 1.1$, implying that the CPU time $T$ grows almost linearly  with
$N$, $T \propto N^{1.1}$.

We conclude that the Monte Carlo unraveling can be implemented for initial
superposition states that are composed of a large number of momentum
eigenstates (say, on the order of $10^2$ to $10^3$). This implies that one may
choose even well localized initial states and consider scenarios where a particle crosses
a slit or a grid. The following sections present numerical results obtained
with such kinds of states. 

\section{Simulation results \label{sec:Numericalresults}}

We proceed to apply the stochastic algorithm to two different types of scattering interactions with the surrounding gas particles. Specifically, we consider the simplest possible scattering process (s-wave hard-sphere scattering) as well as the
case of a general potential, which is treated exactly through partial wave decomposition. Having discussed the determination of the scattering amplitudes, we start out with the simulation of short-time effects. At first, the loss of coherence of an initial superposition of two momentum eigenstates is measured, followed by the treatment of  superpositions of {{spatially}} localized wave packets. 
The latter permits in particular to extract the localization rate discussed in Sect.~\ref{sec:PCD}. As a further example of a decoherence process, counter-propagating localized initial states are considered which lead to the formation of interference patterns. In the course of the evolution, fringe visibility is lost, so that the interplay between coherence and decoherence can be demonstrated.

We then  discuss long-time effects which
exhibit a classical counterpart, starting with energy
relaxation and the approach to thermal equilibrium. Then, 
the spread in position of initially
spatially localized  states is measured, allowing us to observe a transition from quantum dispersion to classical diffusion. 

As discussed in Sect.~\ref{sec:Limitingforms}, the QLBE has several limiting
forms for some of which analytical solutions are known. This permits to demonstrate the validity of the numerical results and to verify the limiting
procedures discussed in \cite{Vacchini2009a}. Further simulations
correspond to situations where the full QLBE is
required. This way physical regimes are entered which have not been
accessible so far, such as decoherence phenomena where the mass of the test particle is comparable to the mass of the gas particles.

\subsection{Scattering amplitudes \label{sec:SA}}

\subsubsection{S-wave hard-sphere scattering \label{sec:SWHS}}

In s-wave hard-sphere scattering the particles are
assumed to be hard spheres with radius $R$, and the kinetic energy to be sufficiently small, $pR \ll \hbar$, such
that only the lowest partial wave
contributes.
In this case the scattering amplitude is independent of the  scattering
angle and the kinetic energy, $
  \left| f \left( \cos \theta ; E_{\tmop{kin}} \right) \right|^2  =  R^2$.
For a constant cross section one can do the
${\tmmathbf{k}_{\bot}}$-integration in the QLBE (\ref{eq:QLBE}). The equation then coincides with the QLBE in Born approximation (\ref{eq:BornD}), such that the numerical results with this interaction should  agree with the stochastic algorithm of Breuer and Vacchini {\cite{Breuer2007a}}. In the  s-wave  examples presented below the system of units is 
defined by setting $\hbar = 1$, $M = 1$ and $R = 1$; the temperature is 
chosen to be $k_B T = 1$ and the gas density is set to one, $n_{\tmop{gas}} = 1$.

An important ingredient for implementing the Monte Carlo unraveling
is the jump rate $\Gamma \left( U \right)$ presented in Eq.~(\ref{eq:QLBG3}).
It is obtained numerically by Monte Carlo integration with
importance sampling (\ref{eq:Importancesample}), based on $n = 10^4$ steps. We find that the collision rate
grows linearly for large momenta, while it saturates  for vanishing $U$ at a value close to
\begin{eqnarray}
  \Gamma_0 & = & n_{\tmop{gas}} v_{\beta} 4 \pi R^2,  \label{eq:SWHS5}
\end{eqnarray}
in agreement  with the analytical prediction in {\cite{Breuer2007a}}.

\subsubsection{Gaussian interaction potential \label{sec:GP}}

Generic scattering processes are characterized by many partial waves with energy dependent scattering phases.
To illustrate the treatment of this general scattering situation, we choose the scattering amplitudes defined by an attractive Gaussian interaction potential
\begin{eqnarray}
  V \left( r \right) & = & -V_0 \, \exp \left( - \frac{r^2}{2 d^2} \right) \, .
  \label{eq:Gauss1}
\end{eqnarray}
\begin{figure}[tb]
  \begin{center}
    \resizebox{12cm}{!}{\includegraphics{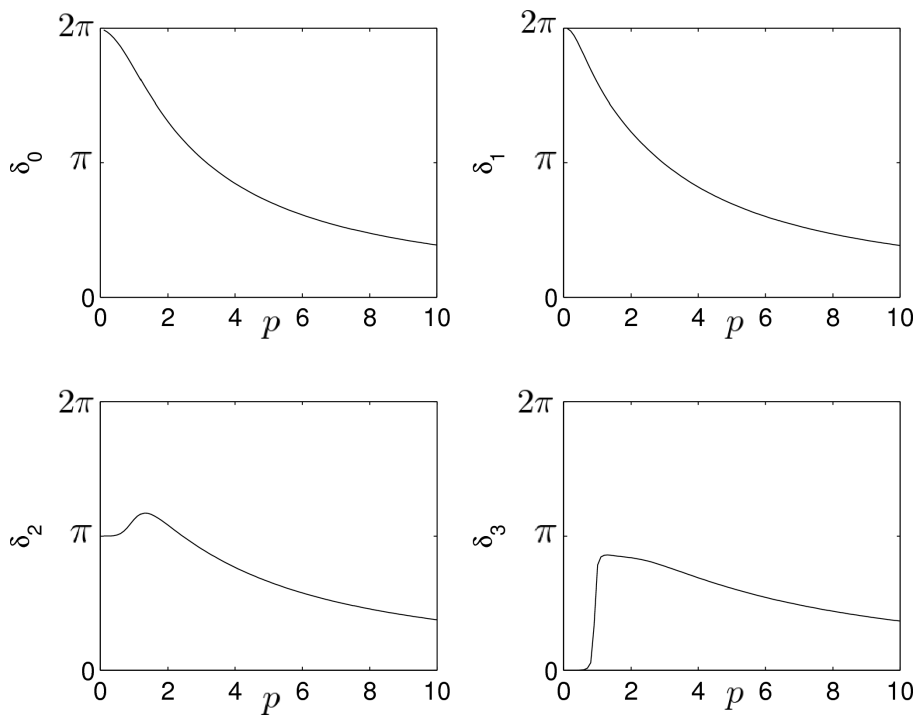}}
  \end{center}
  \caption{The first four phase shifts $\delta_l$ as a function of the
  relative momentum $p$ for the interaction potential (\ref{eq:Gauss1}) and mass ratio $m / M = 1$.  Their asymptotic behavior is in agreement with (\ref{eq:Gauss6}). (There are two bound states for $l = 0, 1$, one for $l = 2$, and no bound state for $l = 3$.) \label{fig:delta}}
\end{figure}
The corresponding scattering amplitude in Born approximation is obtained from (\ref{eq:Bornf1}), which yields
\begin{eqnarray}
  f_B \left( \cos \theta,p \right) & = &  \sqrt{\frac{\pi}{2}} \frac{2
  m_{\ast} V_0 d^3}{\hbar^2} \exp \left( - \frac{d^2 p^2}{\hbar^2} \left[ 1 -
  \cos \theta \right] \right) \, .  \label{eq:Gauss3}
\end{eqnarray}
While this approximation is reliable only for weak interaction potentials, $V_0
\ll E_{\tmop{kin}}$, the exact scattering amplitudes require the energy dependent partial scattering amplitudes
$f_l$ {\cite{Taylor1972a}},
\begin{eqnarray}
  f \left( p, \cos \theta \right) & = & \sum_{l = 1}^{\infty} \left( 2 l + 1
  \right) f_l \left( p \right) P_l \left( \cos \theta \right) \,, \bignone
  \label{eq:Gauss4}
\end{eqnarray}
with $P_l$ the Legendre-polynomials. 

The $  f_l \left( p \right)  =  ({\hbar}/{p}) \exp({i \delta_l}) \sin( \delta_l) $ are related to the partial wave phase shifts $\delta_l$, which can be computed numerically by means of the {\tmem{Johnson algorithm}} {\cite{Johnson1973a}} for a given interaction potential.
If the kinetic energy is large compared to the potential, $V \ll p^2 / 2 M$, the
partial waves are hardly affected by the collision, so that the scattering
amplitudes and phases vanish, $\delta_l \left( p \rightarrow \infty \right) =
0$. For small energies, on the other hand, they behave as {\cite{Taylor1972a}}
\begin{eqnarray}
  \delta_l \left( p \right) & \sim & n_l \pi - a_l p^{2 l + 1} \,,
  \hspace{1em} \tmop{for} \: p \rightarrow 0 \,,  \label{eq:Gauss6}
\end{eqnarray}
with  $a_l$ the scattering lengths and $n_l\in\mathbb{N}_0$. According to the {\tmem{Levinson theorem}}
{\cite{Taylor1972a}}, the integer $n_l$ equals the number of bound states with
angular momentum $l$.
Figure \ref{fig:delta} shows the first four phase shifts for $V_0 =  20$, $d = 1$, and $\hbar = 1$, in agreement  with the Levinson theorem (\ref{eq:Gauss6}) and with
the expected high energy limit. 

For the simulations presented below it is sufficient to include the first $30$ partial waves when evaluating the scattering amplitudes (\ref{eq:Gauss4}). In particular, this ensures that the
optical theorem is satisfied \cite{Taylor1972a}.

\begin{figure}[tb]
  \resizebox{0.7\textwidth}{!}{\includegraphics{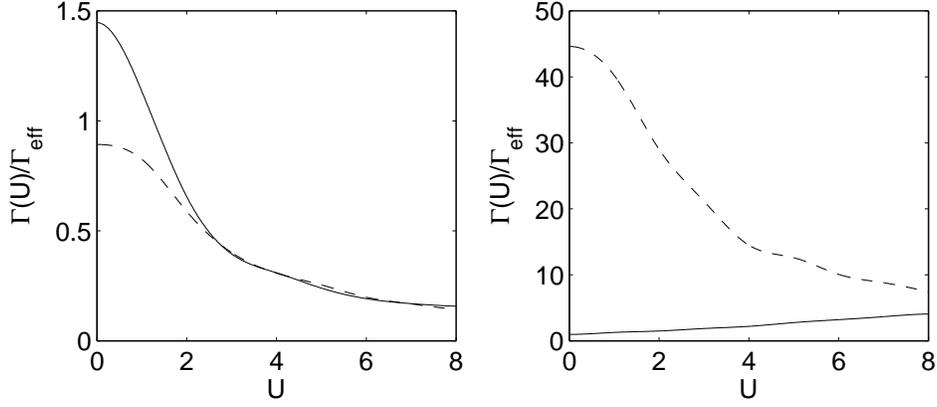}}
  \caption{Jump rate $\Gamma$ as a function of the momentum $U$ assuming the
  Gaussian interaction potential with $V_0 =  1$ (left) and $V_0 =  20$
  (right). The solid line corresponds to the exact scattering amplitude, while the
  dashed line gives the Born approximation. As one expects, the results deviate strongly for large interaction
  potentials. \label{fig:GammaGauss}}
\end{figure}

Figure~\ref{fig:GammaGauss} shows the numerically evaluated jump rate $\Gamma \left( U \right)$ for the Gaussian interaction potential. It is
obtained by a Monte Carlo integration of Eq.~(\ref{eq:QLBG3}) with
importance sampling with $n = 10^4$ steps. The jump rate is given in units of the  collision
rate as defined by the thermal average (\ref{eq:Geff}). The simulation shown
by the solid line in Fig.~\ref{fig:GammaGauss} is based on the exact
scattering amplitude, while the dashed line corresponds to its Born approximation. One observes that the two results differ drastically, in particular for
large interaction potentials $V_0$, while they tend to agree for large momenta $p$,
where the Born approximation is more reasonable.

The Gaussian interaction potential is applied in several examples below. In
these cases, the system of units is defined by setting $\hbar = 1$, $m = 1$
and $d = 1$; moreover, we chose $k_B T = 1$ for the temperature of the gas
environment and the gas density is set to unity, $n_{\tmop{gas}} = 1$.

\subsection{Decoherence in momentum space \label{sec:DM}}

We now apply the Monte Carlo algorithm to the analysis of decoherence effects in momentum space. For this purpose the initial state is taken to be a superposition of two momentum eigenstates,
\begin{eqnarray}
  | \psi \left( 0 \right) \rangle & = &  \alpha \left( 0 \right)
  |\tmmathbf{U} \left( 0 \right) \rangle + \beta \left( 0 \right)
  |\tmmathbf{V} \left( 0 \right) \rangle  \,, 
  \label{eq:DM1}
\end{eqnarray}
which are assumed to have the form
$ \tmmathbf{U} \left( 0 \right) = -\tmmathbf{V} \left( 0 \right) =
  \left( U_0, 0, 0 \right)$. 

Since the states $|\tmmathbf{U} \left( 0 \right) \rangle$ and $|\tmmathbf{V}
\left( 0 \right) \rangle$ are genuine momentum eigenstates, any collision
necessarily leads to an orthogonal state. It follows that the coherences are expected to decay
exponentially
\begin{eqnarray}
  \frac{\left| \langle \tmmathbf{U} \left( 0 \right) | \rho \left( t \right)
  |\tmmathbf{V} \left( 0 \right) \rangle \right|}{\left| \langle \tmmathbf{U}
  \left( 0 \right) | \rho \left( 0 \right) |\tmmathbf{V} \left( 0 \right)
  \rangle \right|} & = & e^{- \Gamma \left( U_0 \right) t} \,,  \label{eq:DM2}
\end{eqnarray}
with the decay rate given by the total collision rate $\Gamma \left( U_0
\right)$.

\begin{figure}[tb]
  \resizebox{0.7\textwidth}{!}{\includegraphics{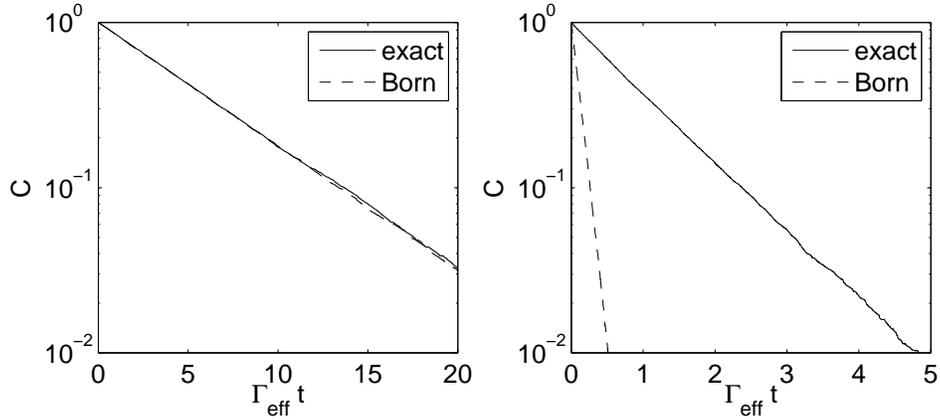}}
  \caption{Semi-logarithmic plot of the ``coherence'' $C \left( t \right)$
  defined in (\ref{eq:DM5}) for the state (\ref{eq:DM1}) with $U_0 =
  \sqrt{6}$. The interaction is described by the Gaussian potential
  (\ref{eq:Gauss1}) with $V_0 =  1$ (left) and $V_0 =  20$ (right). The
  solid line is obtained using the exact scattering amplitude, while the
  dashed line corresponds to the Born approximation. The predictions of the
  decoherence rates differ substantially in case of the large interaction
  potential with $V_0 =  20$. \label{fig:DMgauss}}
\end{figure}

Alternatively, one may view the states $|\tmmathbf{U} \left( 0 \right)
\rangle$, $|\tmmathbf{V} \left( 0 \right) \rangle$ as representing states
which are well localized in momentum space, but with a finite width greater
than the typical momentum transfer. Here a suitable measure for the degree of
coherence is the ensemble average of the coherences exhibited by the
individual quantum trajectories $| \psi \left( t \right) \rangle$
{\cite{Breuer2007a}}, that is
\begin{eqnarray}
  C \left( t \right) & = & \mathbb{E} \left[ \frac{\left| \langle
  \tmmathbf{U} \left( t \right) | \psi \left( t \right) \rangle \langle \psi
  \left( t \right) |\tmmathbf{V} \left( t \right) \rangle \right|}{\left|
  \langle \tmmathbf{U} \left( 0 \right) | \rho \left( 0 \right) |\tmmathbf{V}
  \left( 0 \right) \rangle \right|} \right] \, .  \label{eq:DM3}
\end{eqnarray}
To evaluate this term, recall that the quantum trajectories remain in a
superposition of two momentum eigenstates all the time, so that
$| \psi \left( t \right) \rangle$ has the form
$
  | \psi \left( t \right) \rangle  =  \alpha \left( t \right)
  |\tmmathbf{U} \left( t \right) \rangle + \beta \left( t \right)
  |\tmmathbf{V} \left( t \right) \rangle.
$
By inserting this expression into equation (\ref{eq:DM3}), one finds
{\cite{Breuer2007a}}
\begin{eqnarray}
  C \left( t \right) & = & 2\mathbb{E} \left[ \left| \alpha \left( t \right)
  \beta^* \left( t \right) \right| \right] \, .  \label{eq:DM5}
\end{eqnarray}

Figure \ref{fig:DMgauss} shows a semi-logarithmic plot of the ``coherence'' $C
\left( t \right)$ for the Gaussian interaction potential, choosing an initial momentum
$U_0 = \sqrt{6}$, equal amplitudes $\alpha \left( 0 \right) = \beta \left( 0
\right) = 1 / \sqrt{2}$, and the mass ratio $M / m = 1$. The left-hand side represents a weak interaction potential, and the right-hand side a strong one. In the latter case, the result obtained with the exact
scattering amplitude (solid line) differs markedly from the corresponding Born
approximation (dashed line). The simulation is based on $5 \times 10^3$
trajectories.

This result shows that the full QLBE (\ref{eq:QLBE}) may lead to physical
predictions which deviate significantly from the ones obtained with the QLBE
in Born approximation (\ref{eq:BornD}) if the interaction potential is
sufficiently strong. A similar conclusion is drawn below, when studying
relaxation rates.

The design of experimental tests for decoherence effects in momentum space is
a challenging task {\cite{Vacchini2009a,Rubenstein1999a,Rubenstein1999b}}.
Such a setup would have to provide a source of states with
momentum coherences (as in non-stationary
beams), and it would require an interferometric measurement
apparatus able to detect these coherences. A further difficulty lies in the
inevitable presence and dominance of position decoherence. During the free
evolution a superposition state characterized by two different momentum values will evolve into a
superposition of spatially separated wave packets, which is affected by
decoherence mechanisms in position space {\cite{Vacchini2009a}}.

Position decoherence, in contrast, has already been observed experimentally
in fullerene interference experiments {\cite{hornberger2003a}}. The
following section therefore focuses on the prediction of spatial
decoherence effects based on the Monte Carlo unraveling of the QLBE.

\subsection{Decoherence in position space \label{sec:DIP}}

\subsubsection{Measuring spatial coherences \label{sec:MSC} }

In order to quantify the loss of spatial coherences, i.e. the off-diagonal elements in position representation, $\rho \left( \tmmathbf{X},
\tmmathbf{X}' \right) \equiv \langle \tmmathbf{X}| \rho |\tmmathbf{X}'
\rangle$ one must  assess $\rho \left(
\tmmathbf{X}, \tmmathbf{X}' \right)$ given the quantum trajectories in
the momentum representation, $| \psi \left( t \right) \rangle = \sum_{j = 1}^N
\alpha_j \left( t \right) |\tmmathbf{U}_j \left( t \right) \rangle$. For this
purpose, it is convenient to express the position variable $\tmmathbf{X}$ in
units of the thermal wavelength $\Lambda_{\tmop{th}} = \sqrt{2 \pi \hbar^2 /
mk_B T} \,$,
\begin{eqnarray}
  \tmmathbf{S} & \equiv & \frac{\tmmathbf{X}}{\Lambda_{\tmop{th}}} \, .
  \label{eq:DP0}
\end{eqnarray}
The spatial coherences are then obtained by taking the ensemble average of the
coherences of the individual quantum trajectories, that is
\begin{eqnarray}
  \text{$\rho \left( \tmmathbf{S}, \tmmathbf{S}', t \right)$} & = &
  \mathbb{E} \left[ \langle \tmmathbf{S}| \psi \left( t \right) \rangle
  \langle \psi \left( t \right) |\tmmathbf{S}' \rangle \right] \, .
  \label{eq:DP1a}
\end{eqnarray}
By inserting the momentum representation of $| \psi \left( t \right) \rangle$
into this expression, we find 
\begin{eqnarray}
  &  & \rho \left( \tmmathbf{S}, \tmmathbf{S}', t \right)  \label{eq:DP3}\\
  &  & = \, \frac{1}{\left( 2 \pi \right)^3} \sum_{j, k}^N \mathbb{E} \left[
  \alpha_j \left( t \right) \alpha_k^{\ast} \left( t \right) \exp \left(
  \frac{i}{\hbar} Mv_{\beta} \Lambda_{\tmop{th}} \, \left[ \tmmathbf{S} \cdot
  \tmmathbf{U}_j \left( t \right) -\tmmathbf{S}' \cdot \tmmathbf{U}_k \left( t
  \right) \right] \right) \right] \bignone \, , \nonumber
\end{eqnarray}
which allows us to compute the time evolution of the coherences (\ref{eq:DP1a}) by means of the amplitudes $\alpha_j \left( t
\right)$ and the scaled momenta $\tmmathbf{U}_j \left( t \right)$.

\begin{figure}[tb]
  \resizebox{.7\textwidth}{!}{\includegraphics{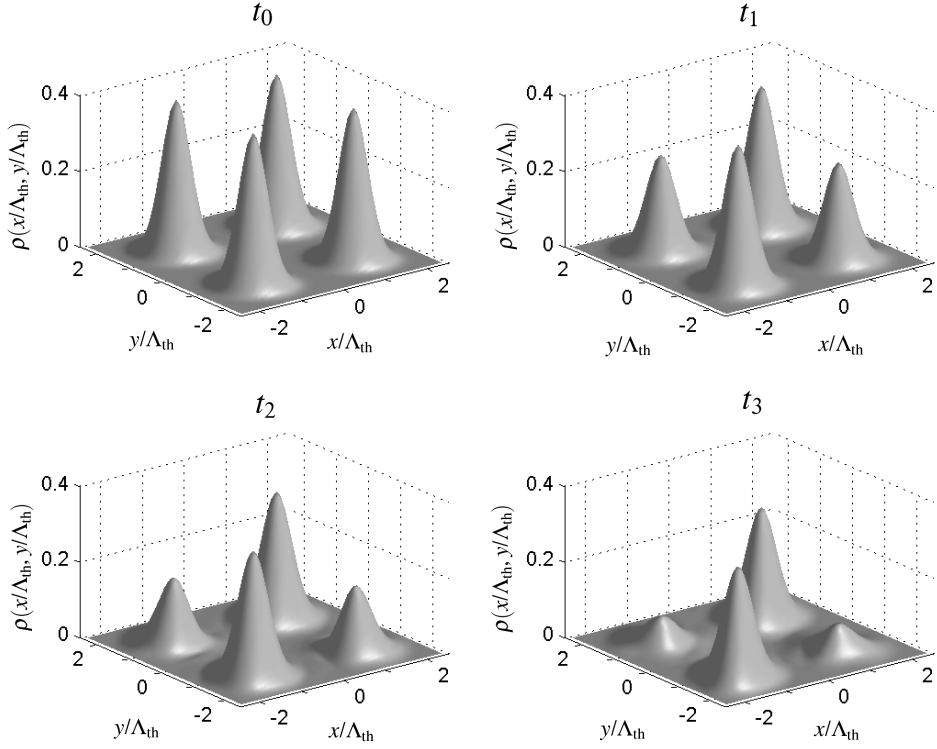}}
  \caption{Evolution of the density matrix in the position representation for an
  initial superposition of two Gaussian wave packets, obtained by solving the
  three-dimensional QLBE for s-wave hard-sphere scattering. The spatial
  coherences $\rho \left( x / \Lambda_{\tmop{th}}, y / \Lambda_{\tmop{th}} \right)$ are expressed in
  units of the thermal wavelength.
  \label{fig:FDM}}
\end{figure}

A typical application might describe a particle passing through an interferometer, where it is spatially localized in one spatial direction, and is characterized by an incoherent  distribution of momenta in the
other two directions. From now on, we therefore restrict the discussion  to
initial states of the form
\begin{eqnarray}
  | \psi \left( 0 \right) \rangle & = & \sum_{j = 1}^N \alpha_j \left( 0
  \right) |U_j \left( 0 \right), V \left( 0 \right), W \left( 0 \right)
  \rangle \bignone \, \,,  \label{eq:DP4}
\end{eqnarray}
where $|U_j \left( 0 \right), V \left( 0 \right), W \left( 0 \right) \rangle$
denote scaled eigenstates of the momentum operator $\mathsf{P} \equiv \left(
\mathsf{P}_x, \mathsf{P}_y, \mathsf{P}_z \right)$. By taking $N$ to be
sufficiently large, Eq.~(\ref{eq:DP4}) may represent states which are localized in
one spatial direction. Due to the conservation of momentum superpositions, the
ensuing quantum trajectories have the structure
\begin{eqnarray}
  | \psi \left( t \right) \rangle & = & \sum_{j = 1}^N \alpha_j \left( t
  \right) |U_j \left( t \right), V \left( t \right), W \left( t \right)
  \rangle \, .  \label{eq:DP5}
\end{eqnarray}
The assessment of spatial coherences (\ref{eq:DP3}) can be simplified in this case
by focusing on the coherences in $x$-direction,
\begin{eqnarray}
  &  & \rho \left( \left[ S, 0, 0 \right], \left[ S', 0, 0 \right], t \right)
  \label{eq:DP6}\\
  &  & = \, \frac{1}{\left( 2 \pi \right)^3} \sum_{j, k}^N \mathbb{E} \left[
  \alpha_j \left( t \right) \alpha_k^{\ast} \left( t \right) \exp \left(
  \frac{i}{\hbar} Mv_{\beta} \, \Lambda_{\tmop{th}} \left[ SU_j \left( t
  \right) - S' U_k \left( t \right) \right] \right) \right] \, . \nonumber
\end{eqnarray}

To visualize the evolution of the density matrix in position representation, we
consider an initial superposition of two resting Gaussian wave packets, with
scaled mean positions $\langle \mathsf{S} \rangle_{1, 2} = \pm 1.2$ and width
$\sigma_{1, 2} = 0.2$ (in units of $\Lambda_{\tmop{th}}$). This state may be
written in the form (\ref{eq:DP5}) by using a finite-dimensional
representation of the corresponding Fourier transform. Figure \ref{fig:FDM}
depicts the ensuing evolution of the matrix elements (\ref{eq:DP6}), obtained
by solving the QLBE under the assumption of s-wave hard-sphere scattering and
equal masses $m = M$. It shows four snapshots of the density matrix for the
scaled times ${t} \Gamma_0 \in\{ 0, 1 / 3, 2 / 3, 4 / 3 \}$.
The simulation is based on $10^3$ realizations of the stochastic process and
the state is represented using $N=55$ momentum eigenstates. 

\subsubsection{Measuring the localization rate \label{sec:MLF}}

\begin{figure}[tb]
  \resizebox{0.8\textwidth}{!}{\includegraphics{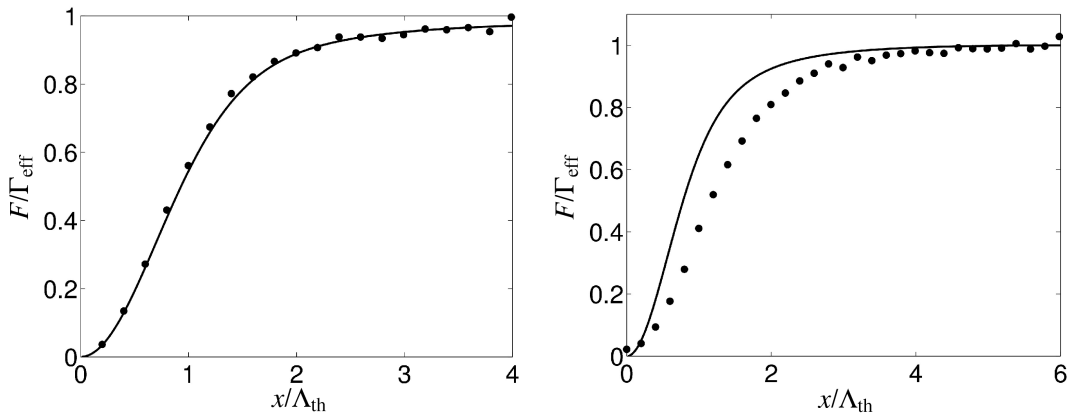}}
  \caption{Decay rate of the spatial coherences as a function of the wave packet 
  separation for a
  Gaussian interaction potential and the mass ratios $M / m = 100$ (left) and
  $m = M$ (right). 
  The solid line shows the prediction of pure collisional
  decoherence, Eq.~(\ref{eq:explicitformloc}), and the dots give the
  result of the stochastic simulation of the QLBE. One observes that the
  predictions of the two models agree for $M \gg m$, while they deviate for
  $m = M$. The localization rate saturates in all cases at the average
  collision rate $\Gamma_{\tmop{eff}}$. We note that the decay rate of the QLBE does not vanish for $m = M$ as $x \to 0$, since there is a loss of the
  populations due to diffusion. \label{fig:Locgauss} }
\end{figure}

As discussed in Sect.~\ref{sec:PCD}, the QLBE simplifies to the master equation of pure collisional decoherence if one assumes the tracer particle to be much heavier than the
gas particles. In this model the decay rate $F$ of spatial coherences is a
function of the distance $x=|\tmmathbf{X}-\tmmathbf{X}'|$ only; it does
not depend on the particular matrix elements of the state, see
Eq.~(\ref{eq:explicitformloc}). Hence, the decoherence process is completely
characterized by the localization rate $F \left(x\right)$.

By evaluating the decoherence rates for various mass ratios and initial
states, we found that this behavior is observed even in regimes where the QLBE
does not reduce to the master equation of collisional decoherence. This suggests that the
decoherence dynamics of the QLBE is generally characterized by a
one-dimensional function $F \left(x\right)$. 

Figure \ref{fig:Locgauss} shows the localization rate for the Gaussian
interaction potential with $V_0 =  1$ and the mass ratios $M / m = 100$
(left) and $M / m = 1$ (right). The dots give the decay rate as
evaluated from (\ref{eq:DP6}), obtained by $\text{$5 \times 10^4$}$
realizations of the Monte Carlo unraveling of the QLBE. The solid line
represents the localization rate of collisional decoherence
(\ref{eq:explicitformloc}), calculated by numerical integration.
As expected, one finds an excellent agreement between the predictions of
collisional decoherence and the solution of the QLBE if the test particle mass
is much larger than the gas mass, $M / m = 100$. 

Moreover, it turns out that the results of the two models do not differ
substantially even for equal masses $m = M$. This holds in particular for
large distances, where the decay rates converge to the average collision rate
$\Gamma_{\tmop{eff}}$ (in all cases). Indeed, in this limit one collision should be sufficient to
reveal the full `which path' information, so that a saturation at
$\Gamma_{\tmop{eff}}$ is expected. For equal masses the prediction of the QLBE
does not tend to zero in the limit of small distances, $F \left( 0 \right) >
0$. This is due to the contribution of quantum diffusion, which is 
more pronounced when the test particle is lighter. 

\subsection{Interference and decoherence \label{sec:ID}}

\begin{figure}[tb]
  \resizebox{0.7\textwidth}{!}{\includegraphics{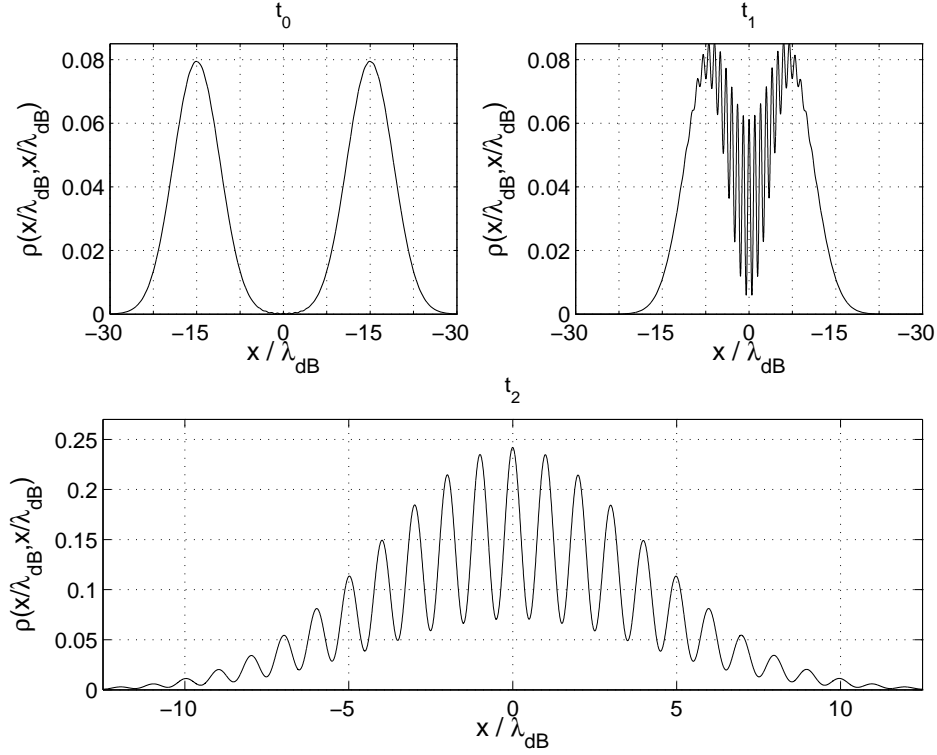}}
  \caption{Evolution of the position diagonal elements of the density matrix $\rho \left( x /
  \lambda_{\tmop{dB}}, x / \lambda_{\tmop{dB}} \right)$ for an initial
  superposition of two counter-propagating minimum-uncertainty wave packets. 
  The figure, obtained by
  solving the QLBE for s-wave hard-sphere scattering, shows three
  snapshots of the dynamics at times $\Gamma_0 \left( t_0, t_1, t_2 \right) =
  \left( 0, 9, 18 \right)$. The increasing influence of decoherence manifests itself as a reduced visibility of the interference fringes as time progresses. \label{fig:Interference}}
\end{figure}

To illustrate the
interplay between coherent and incoherent dynamics, let us study how the
formation of interference patterns is affected by the interaction with the
background gas. To this end, consider the scenario depicted in
Fig.~\ref{fig:Interference}. Here the $x$-component of the three-dimensional
initial state is prepared in a superposition of two counter-propagating
minimum-uncertainty wave packets $\psi_{1, 2}$, while the other two components have a definite
momentum. The wave packets start overlapping in the course of the evolution, and 
their interference leads to oscillations of the spatial
probability density $\rho \left( x, x, t \right)$, with a period given by the de Broglie wavelength $\lambda_{\tmop{dB}}$
associated to the relative momentum between the minimum-uncertainty wave packets. Besides this
coherent effect, one observes an increasing signature of decoherence,  the gradual loss of fringe visibility; this becomes evident in particular
in the bottom  panel of Fig.~\ref{fig:Interference}.

Figure \ref{fig:Interference} is obtained by the Monte Carlo unraveling of the
QLBE, assuming s-wave hard-sphere scattering and a mass ratio $M / m = 100$.
The parameters of the simulation are conveniently expressed in terms of the
de Broglie wavelength $\lambda_{\tmop{dB}}$ and the scattering rate $\Gamma_0$
(\ref{eq:SWHS5}), which serve to define the dimensionless variables
\begin{eqnarray}
  &  & S_{\tmop{dB}} \equiv \frac{X}{\lambda_{\tmop{dB}}} \,, \hspace{2em}
  U_{\tmop{dB}} \equiv \frac{P}{M \lambda_{\tmop{dB}} \Gamma_0} \, .
  \label{eq:IDS1}
\end{eqnarray}
In this system of units the position and momentum expectation values of the coherent
states $\psi_{1, 2}$ read as $\langle \mathsf{S}_{\tmop{dB}} \rangle_{1, 2} =
\mp 15$ and $\langle \mathsf{U}_{\tmop{dB}} \rangle_{1, 2} = \pm 0.9$; their
width is characterized by the standard deviation $\sigma_{1, 2} /
\lambda_{\tmop{dB}} = 4$, and the de Broglie wavelength is fixed by
setting $\lambda_{\tmop{dB}} / \Lambda_{\tmop{th}} = 2.5 \times 10^{- 2}$.
The figure shows three snapshots of the populations of the density matrix for the
scaled times $\Gamma_0 \left( t_0, t_1, t_2 \right) = \left( 0, 9, 18
\right)$. The simulation is based on $\text{$2.5 \times 10^4$}$ realizations
of the stochastic process.

As mentioned above, an important quantity to 
characterize the loss of quantum coherence is the fringe visibility, which we define here as
the difference between the central maximum and the neighboring minimum divided by
their sum. In the last snapshot, shown at the bottom of Fig.~\ref{fig:Interference}, one extracts a visibility of $\mathcal{V}=55\%$. To understand this result quantitatively, let us estimate the decay rate of
the visibility by means of the integrated localization rate,
\begin{eqnarray}
  \mathcal{V} \left( \tau \right) & = & \exp \left( - \int_0^{\tau} \mathd
  \tau' \bignone F \left[ S \left( \tau' \right) \right] \tau' \right)
  \mathcal{V}  \left( 0 \right) \,,  \label{eq:ID2}
\end{eqnarray}
where $S \left( \tau \right) = \left| \langle \mathsf{X} \left( \tau \right)
\rangle_1 - \langle \mathsf{X} \left( \tau \right) \rangle_2 \right| /
\Lambda_{\tmop{th}}$ denotes the distance of the minimum-uncertainty wave packets in units of
the thermal wavelength at time $\tau$. By noting that the wave packets move in
absence of an external potential, one finds
\begin{eqnarray}
  S \left( \tau \right) & = & \frac{\lambda_{\tmop{dB}}}{\Lambda_{\tmop{th}}}
  \left( \left| \langle \mathsf{S}_{\tmop{dB}} \rangle_1 - \langle
  \mathsf{S}_{\tmop{dB}} \rangle_2 \right| - \tau \left| \langle
  \mathsf{U}_{\tmop{dB}} \rangle_1 - \langle \mathsf{U}_{\tmop{dB}} \rangle_2
  \right| \right) \, .  \label{eq:ID5}
\end{eqnarray}
Since the tracer particle is much heavier than the gas molecules
the dynamics described by the QLBE should be well approximated by the master
equation (\ref{eq:Collisionaldecoherence2}) of pure collisional decoherence. In this case $F$ is described by the formula (\ref{eq:explicitformloc}), which can be
evaluated analytically in the case of s-wave hard-sphere scattering,
\begin{eqnarray}
  F \left( S \right) & = & 2 \sqrt{\pi} n_{\tmop{gas}} R^2 v_{\beta} \left[ 4
  - S^{- 1} \exp \left( - 4 \pi S^2 \right) \tmop{erfi} \left( 2 \sqrt{\pi} S
  \right) \right] \,,  \label{eq:SWHS4}
\end{eqnarray}
where $\tmop{erfi} \left( x \right) \equiv - i \tmop{erf} \left( ix \right)$
denotes the imaginary error function.

A prediction for the visibility (\ref{eq:ID2}) may then be obtained by a simple numerical integration. This  yields $\mathcal{V} \left( t_2 \right)  \simeq  56\%$, in good agreement with the value of $\mathcal{V}=55\%$ obtained by the stochastic solution of the full QLBE.

\subsection{Relaxation and thermalization \label{sec:Rel}}

We now study the long-time behavior of the energy
expectation value.  As discussed in \cite{Vacchini2009a} any solution of the QLBE will approach the canonical thermal state  asymptotically. 
The kinetic energy in the simulation must therefore  converge to the thermal energy $3 / \left(
2 \beta \right)$. Expressed in dimensionless units, this means that
{\cite{Vacchini2009a}}
\begin{eqnarray}
  \langle \mathsf{U}^2 \rangle_t & \rightarrow & \langle \mathsf{U}^2
  \rangle_{\tmop{eq}} = \frac{3}{2} \frac{m}{M} \,, \hspace{2em} \tmop{for} \:
  t \gg \gamma \,^{- 1},  \label{eq:R1}
\end{eqnarray}
with $\gamma$ the relaxation rate.

\begin{figure}[tb]
  \begin{center}
    \resizebox{0.6\textwidth}{!}{\includegraphics{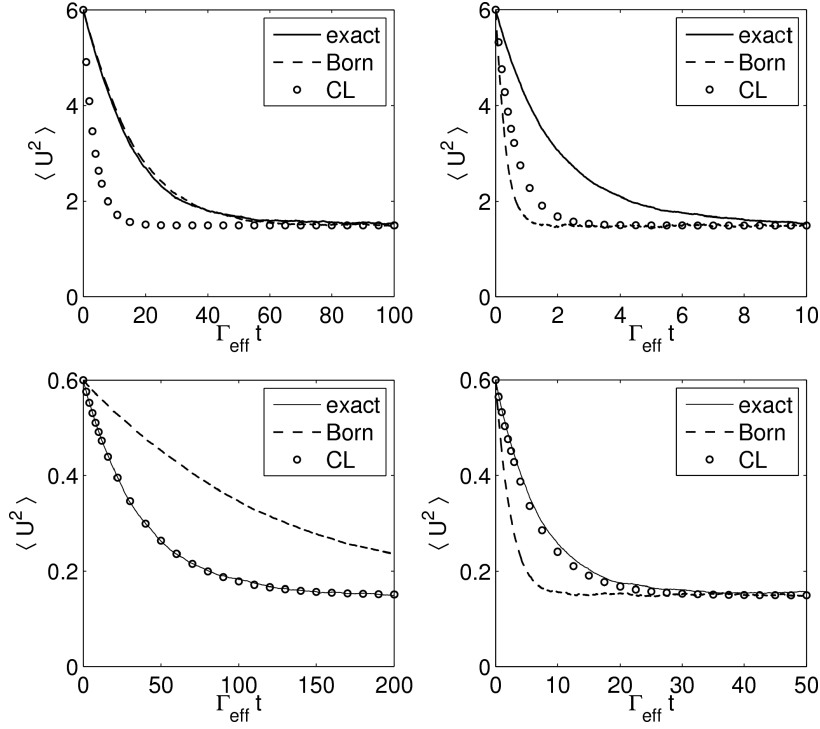}}
  \end{center}
  \caption{Energy relaxation as obtained by using the exact scattering amplitude of the
  Gaussian interaction potential, compared to the corresponding Born approximation and
  the solution of the CL equation (\ref{eq:relaxationrate}). We choose the
  potential strengths $V_0 =  1$ (left) and $V_0 =  20$ (right) and the mass
  ratios $M / m = 1$ (top) and $M / m = 10$ (bottom). The correct
  equilibrium values (\ref{eq:R1}) of $3 / 2$ (top) and $3 / 20$ (bottom) are obtained 
  in all cases. One observes that the
  exact result agrees with the solution of the CL equation for a heavy tracer
  particle (bottom), while it deviates strongly for $M=m$ (top). The Born approximation
  gives reliable results only when the
  kinetic energy is much greater than the potential one (top left), which indicates that it may
  underestimate the energy relaxation even for weak interactions 
  if the test particle has a large mass (bottom left).
  \label{fig:Relgauss}}
\end{figure}

If the state is
close to thermal,  and if the tracer particle is much heavier than the gas
particles the QLBE reduces to the
Caldeira-Leggett (CL) equation in Lindblad form (\ref{eq:CL}). The corresponding  energy behavior is then well understood {\cite{Breuer2007b,Vacchini2009a}}.
\begin{eqnarray}
  \langle \mathsf{U}^2 \rangle^{}_t & = & \langle \mathsf{U}^2
  \rangle_{\tmop{eq}} + \left( \langle \mathsf{U}^2 \rangle_{t_0} - \langle
  \mathsf{U}^2 \rangle_{\tmop{eq}} \right) e^{- 4 \gamma_{} t} \,.
  \label{eq:RelEnergy}
\end{eqnarray}
Moreover, the relaxation rate $\gamma$ can be
expressed in terms of the microscopic quantities, see Eq.~(\ref{eq:relaxationrate}). The integral can be evaluated analytically for the case of a constant
cross section {\cite{Vacchini2009a}},
\begin{eqnarray}
  \gamma & = & \frac{4}{3 \sqrt{\pi}} \frac{m}{M} \Gamma_0 \, .
  \label{eq:Gammaconst}
\end{eqnarray}

Figure \ref{fig:Relgauss} shows the energy relaxation exhibited by the
stochastic solution of the QLBE for a weak and a strong Gaussian interaction potential, with
$V_0 =  1$ (left) and $V_0 =  20$ (right) respectively. The solid line depicts the solution
of the QLBE based on the exact scattering amplitude, while the corresponding
Born approximation is represented by the dashed line; both simulations are
based on $5 \times 10^3$ trajectories. The initial state is here a momentum
eigenstate with dimensionless eigenvalue $U_0 = \sqrt{6}$ (top) and $U_0 = \sqrt{0.6}$
(bottom), corresponding to mass ratios of $M/m=1$ and $M/m=10$, respectively. In case of a relatively large tracer mass (bottom) one
obtains a good agreement with the prediction of the CL equation (\ref{eq:RelEnergy}) (open dots). Here the relaxation rate was obtained by
numerical integration of the right-hand side of Eq.~(\ref{eq:relaxationrate}).
For equal masses, on the other hand, the results deviate noticeably (top). As expected,
all of the solutions converge to the correct equilibrium values, given by the
scaled energies  $3 / 2$ (top) and $3 / 20$ (bottom). The Born
approximation yields reliable results only in the situation depicted by the top left panel, where the kinetic energy is much larger
than the potential.

Again, we are led to conclude that the full QLBE (\ref{eq:QLBE}) may
give rise to predictions which deviate significantly from the ones
obtained with the QLBE in Born approximation (\ref{eq:BornD}). This holds in
particular for strong interaction potentials, where the corresponding
scattering amplitudes are different. Furthermore, this result verifies that
the expression (\ref{eq:relaxationrate}) obtained in
{\cite{Hornberger2008a,Vacchini2009a}} yields the correct relaxation rate in
the quantum Brownian limit.

\subsection{Diffusion \label{sec:Diffusion}}

As a final aspect we study the quantum diffusion
process described by the QLBE. To this end, a localized initial state is
prepared and the growth of the position variance is
measured. Before discussing the numerical result, we summarize
analytical predictions based on {\cite{Vacchini2009a}}.

On short time scales, where the number of collisions is small, one expects the
variance growth to be dominated by quantum dispersion. This implies that the
variance growth is parabolic; for an initial state of minimum uncertainty one expects
\begin{eqnarray}
 \sigma^2_X \left(  t \right) & = &\sigma^2_X \left(  0 \right) +
  \frac{\hbar^2}{4 M^2\sigma^2_X \left(  0 \right)} t^2 \,.
  \label{eq:Dis1}
\end{eqnarray}

At time scales after which many collisions have occurred the variance
growth is expected to be dominated by classical diffusion. The
corresponding diffusion constant can be estimated by considering that the QLBE approaches asymptotically the
classical linear Boltzmann equation asymptotically. The latter can be simplified, by
taking the {\tmem{Brownian limit}} of heavy tracer
particles with a momentum $P$ close to the typical thermal value
$P_{\beta} = \sqrt{2 M / \beta}$ 
{\cite{Rayleigh1891a,Green1951a,Vacchini2009a}}. Under these conditions, the
classical linear Boltzmann equation  (\ref{eq:CLBE}) reduces to the Kramers equation
{\cite{Vacchini2009a}}
\begin{eqnarray}
  \partial_t w \left( \tmmathbf{P} \right) & = & \eta \sum_{i = 1}^3 \left(
  \frac{\partial}{\partial P_i} \left[ P_i w \left( \tmmathbf{P} \right)
  \bignone \bignone \right] + \frac{M}{\beta} \frac{\partial^2}{\partial
  P_i^2} w \left( \tmmathbf{P} \right) \right) \,  \label{eq:Dif1}
\end{eqnarray}
for the momentum distribution $w \left( \tmmathbf{P} \right)$, with $\eta$ the friction
coefficient. The latter can be expressed in terms of the microscopic
details of the gas {\cite{Ferrari1987a,Vacchini2009a}}, yielding $\eta
= 2 \gamma$, with $\gamma$ the relaxation rate appearing in the
Caldeira-Leggett equation, see Sect.~\ref{sec:DL},
Eq.~(\ref{eq:relaxationrate}).

The Kramers equation predicts normal diffusion, i.e. a linear growth of the variance 
$ \sigma^2_X \left(  t \right)  = \sigma^2_X \left(  0 \right) + 2 Dt$, with diffusion constant $D = \eta M / \beta$ {\cite{Kampen2006a}}. This
leads to the prediction that
\begin{eqnarray}
 \sigma^2_X \left(  t \right) & = &\sigma^2_X \left(  0 \right) +
  \frac{1}{\beta M \gamma} t \,   \label{eq:Dif2}
\end{eqnarray}
whenever the Brownian limit of the QLBE is applicable.
For the case of a constant cross section, where $\gamma$ can be evaluated analytically
(see Eq.~(\ref{eq:Gammaconst})),  Eq.~(\ref{eq:Dif2}) provides an
analytical prediction for the diffusion constant. It is expected to be valid when $M \gg m$.

\begin{figure}[tb]
  \begin{center}
    \resizebox{0.8\textwidth}{!}{\includegraphics{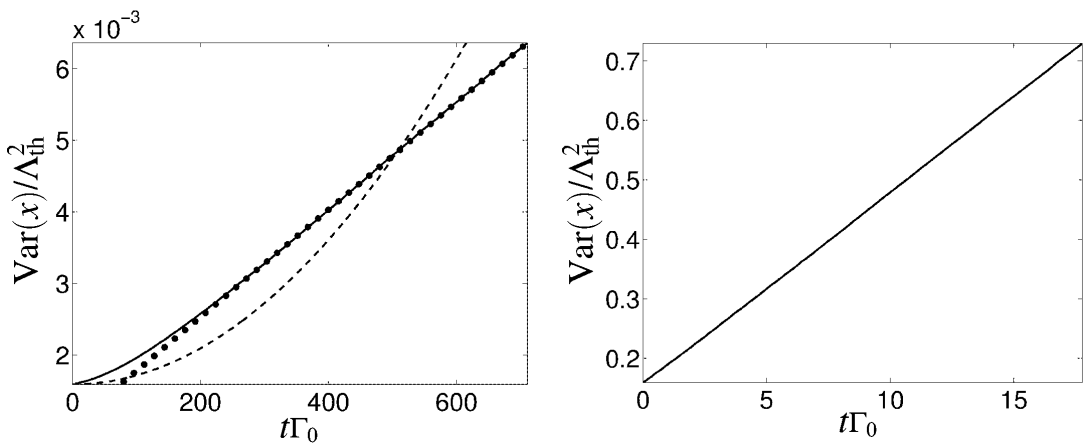}}
  \end{center}
  \caption{The solid lines give the time dependence of the spatial variance,
  as obtained by solving the QLBE for s-wave hard-sphere scattering and mass
  ratios of $M / m = 100$ (left) and $M / m = 1$ (right). Left: after an 
  initial quadratic increase the variance displays the straight line
  behavior expected of classical diffusion. The fit displayed by the
  dotted line has a slope close to the diffusion constant predicted by the
  Kramers equation (relative error: $12\%$). For comparison, the dashed line gives the dispersive broadening of the initial wave packet in absence of a gas. 
  Right: for a test particle with a mass equal to that of the gas particles the
  variance growth is dominated by classical diffusion
  even on the time scale of a single collision. \label{fig:diff}}
\end{figure}

The solid line in Fig.~\ref{fig:diff} shows the variance growth of the spatial
populations, obtained by solving the QLBE for s-wave hard-sphere scattering
and mass ratios $M / m = 100$ (left) and $M / m = 1$ (right). This stochastic
simulation is based on $4 \times 10^3$ trajectories. The initial state is chosen to be a Gaussian with width $\sigma^2_X \left(  0 \right) /
\Lambda_{\tmop{th}}^2 = 1.6 \times 10^{- 3}$ (left) and $\sigma^2_X \left( 
0 \right) / \Lambda_{\tmop{th}}^2 = 1.6 \times 10^{- 1}$ (right).

Let us first focus on the left-hand
side panel which corresponds to a very massive particle. Here the solution of the QLBE starts with a quadratic
dependence at  small times. The curvature is unrelated to that of free quantum dispersion  (dashed line), Eq.~(\ref{eq:Dis1}), which is clearly due to the large number of collisions occurring on the time scale of the wave packet broadening. (Time is given in terms of the average period between collisions in our dimensionless units.)

After a time corresponding to about 200 collisions the curve displays the straight line behavior expected for 
classical diffusion. A linear fit to this straight part, indicated by the
dotted line, has a slope of approximately $7.5 \times 10^{- 6}$. This differs by about $12\%$ from the
analytical considerations presented above, where one expects a straight line of the
form
\begin{eqnarray}
  \frac{\sigma^2_X \left(  t \right)}{\Lambda_{\tmop{th}}^2} & = &
  \frac{\sigma^2_X \left(  0 \right)}{\Lambda_{\tmop{th}}^2} + \left( \beta
  M \gamma \Lambda_{\tmop{th}}^2 \Gamma_0 \right)^{- 1} t \Gamma_0 \nonumber\\
  & = & 1.6 \times 10^{- 3} + 6.7 \times 10^{- 6} \, t \Gamma_0 \, .
  \label{eq:Dif3}
\end{eqnarray}

For equal masses (right panel) one obtains a straight line already starting from small
times, which indicates that classical diffusion dominates over quantum
dispersion. The slope of about $3.2 \times 10^{- 2}$ implies
that the diffusion constant is much greater for light test particles. However, these
results cannot be related to Kramers equation, since the latter is valid only in the  Brownian limit of large tracer masses.

\section{Conclusions \label{sec:Conclusion}}\label{sec:conclusion}

We presented a stochastic algorithm for solving the full quantum linear
Boltzmann equation given an arbitrary interaction potential. By exploiting the
translational invariance of the QLBE it allows one to efficiently propagate
superpositions of momentum eigenstates without increasing 
the dimension of the state space. Since the computation time scales
almost linearly with the number of basis states, arbitrary states can be
represented in practice, in particular spatially localized ones. This enables
us to simulate many important physical processes, ranging from short-time
effects, such as the loss of fringe visibility in interference experiments, to
long-time relaxation and thermalization phenomena.

For the cases of s-wave hard-sphere scattering and a Gaussian interaction potential, we analyzed the range of validity of different
limiting forms of the QLBE, including the collisional decoherence model, the
quantum Brownian limit and the classical linear Boltzmann equation. Moreover,
we compared the solutions of the full QLBE to those of the simplified equation
in Born approximation. Here it is found, for the above interactions, that the full QLBE may lead to physical predictions which deviate significantly from the ones obtained with
the QLBE in Born approximation if the interaction potential is sufficiently
strong.

This method will find applications, e.g.~in describing interference
experiments with species, whose mass is smaller than or comparable to the mass of the gas particles. The existing methods are not able to quantify the loss of coherence in such a situation.
Moreover, future studies might consider extensions of the discussed method to the
recently developed quantum master equation for the collisional dynamics of
particles with internal degrees of freedom 
\nocite{PhysRevA.78.022112}
\nocite{HemmingKrems}
\cite{PhysRevA.78.022112,HemmingKrems,Smirne2010a}. Even though
this equation is more involved than the QLBE, it is also translational
invariant. Since this property is the main prerequisite for the present
algorithm, it should be extensible to the quantum master equation of
{\cite{Smirne2010a}}.

\subsection*{Acknowledgments}
We would like to thank B.~Vacchini for many helpful discussions. This work was partially funded by the DFG Emmy Noether program. M.B. also acknowledges support by the QCCC Program of the Elite Network of Bavaria.

\end{document}